\documentclass[reprint, secnumarabic,amssymb, nobibnotes, aps, prx]{revtex4-2}
\usepackage{amsmath}
\usepackage{amsfonts}
\usepackage{amssymb}
\usepackage{multirow}
\usepackage[hidelinks]{hyperref}
\usepackage{xcolor}
\hypersetup{
    colorlinks,
    linkcolor={red!80!black},
    citecolor={blue!80!black},
    urlcolor={blue!80!black}
}

\usepackage{color}

\usepackage{graphicx}

\usepackage{enumerate}
\usepackage{comment}

\newcommand{\br}[1]{\left( #1 \right)}
\newcommand{\cbr}[1]{\left\{ #1 \right\}}
\newcommand{\sbr}[1]{\left[ #1 \right]}
\newcommand{\abr}[1]{\left\langle #1 \right\rangle}
\newcommand{\pd}[1]{\frac{\partial}{\partial #1}}

\newcommand{\rhobar}{\bar{\rho}}
\newcommand{\intc}{\mathcal{Q}}
\newcommand{\hata}{\hat{a}}

\newcommand{\iu}{\textbf{\textit{i}}}
\newcommand{\jfl}{\mathcal{J}^{(fl)}}
\newcommand{\jd}{\mathcal{J}^{(d)}}
\newcommand{\tp}{t^{\prime}}
\newcommand{\tpp}{t^{\prime \prime}}
\newcommand{\tppp}{t^{\prime \prime \prime}}
\newcommand{\imgi}{\mathrm{\mathbf{i}}}
\newcommand{\actp}{a^\prime}

\newcommand{\be}{\begin{equation}}
\newcommand{\ee}{\end{equation}}
\newcommand{\bea}{\begin{eqnarray}}
\newcommand{\eea}{\end{eqnarray}}

\def\lsim {\protect \raisebox{-0.75ex}[-1.5ex]{$\;\stackrel{<}{\sim}\;$}}

\begin{document}





\title{Anomalous relaxation and hyperuniform fluctuations in center-of-mass conserving systems with broken time-reversal symmetry}

\author{Anirban Mukherjee}
\email{anirban9973@gmail.com}
\author{Dhiraj Tapader}
\author{Animesh Hazra}
\email{animesh\_edu@bose.res.in}
\author{Punyabrata Pradhan}
\affiliation{Department of Physics of Complex Systems, S. N. Bose
   National Centre for Basic Sciences, Block-JD, Sector-III, Salt Lake,
   Kolkata 700106, India.}
\begin{abstract}
  We study a paradigmatic model of absorbing-phase transition - the Oslo model - on a  one-dimensional ring of $L$ sites with a fixed global density $\rhobar$; we consider the system strictly above criticality. Notably, microscopic dynamics conserve both mass and {\it center of mass (CoM), but lacks time-reversal symmetry}.
   We show that, despite having highly constrained dynamics due to CoM conservation, the system exhibits diffusive relaxation away from criticality and superdiffusive relaxation near (above) criticality.
   Furthermore, the CoM conservation severely restricts particle movement, rendering the mobility - a transport coefficient analogous to the conductivity for charged particles - to vanish exactly. Indeed the temporal growth of current fluctuation is qualitatively different from that observed in diffusive systems with a single conservation law. Away from (above) criticality, steady-state fluctuation $\abr{\intc_i^2(T,\Delta)}$ of current $\mathcal{Q}_i$ across $i$th bond up to time $T$ {\it saturates} as  $\abr{\intc_i^2} \simeq \Sigma_Q^2(\Delta) - {\rm const.} T^{-1/2}$; near criticality, it grows subdiffusively as $\abr{\intc_i^2} \sim T^\alpha$, with $0 < \alpha < 1/2$, and eventually {\it saturates} to $\Sigma_Q^2(\Delta)$. The asymptotic current fluctuation $\Sigma_Q^2(\Delta)$ is a {\it nonmonotonic} function of $\Delta$: It diverges as $\Sigma_Q^2(\Delta) \sim \Delta^2$ for $\Delta \gg \rho_c$ and $\Sigma_Q^2(\Delta) \sim \Delta^{-\delta}$, with $\delta > 0$, for $\Delta \to 0^+$. By using a mass-conservation principle, we exactly determine the exponents $\delta = 2(1-1/\nu_{\perp})/\nu_{\perp}$ and $\alpha = \delta/z \nu_{\perp}$ via the correlation-length and dynamic exponents, $\nu_{\perp}$ and $z$, respectively.
    Finally, we show that, in the steady state, the self-diffusion coefficient $\mathcal{D}_s(\rhobar)$ of tagged particles is connected to activity by $\mathcal{D}_s(\rhobar) = a(\rhobar) / \rhobar$.

\end{abstract}

\maketitle


\section{Introduction}

Recently there has been a surge in interest in characterizing transport in quantum many-body systems,  such as fractonic fluids ~\cite{Prem2018Feb, Gromov2020Jul, Feldmeier2020Dec, Morningstar2020Jun, Sala2020Feb, Pai2019Apr, Glorioso2022Aug, Scherg2021Jul, Shenoy2020Feb}, which have more than one conservation law and consequently possess an interesting dynamical structure. Indeed, by using phenomenological hydrodynamic theory, it was claimed that the time evolution of initially localized charge density in dipole-moment conserving systems, with short-ranged interaction, can only have subdiffusive temporal growth  ~\cite{Feldmeier2020Dec, Morningstar2020Jun, Han2023Apr, Glorioso2022Aug}. The class of models studied in these works typically have time-reversal symmetry present in the systems.  
Of course, broken time-reversal symmetry allows for more choices in microscopic dynamics \cite{Voituriez-NJP2019} and its role on the large-scale properties of dipole-moment conserving systems is examined by us in this study. We provide a counter-example of a paradigmatic model system - the celebrated Oslo model  \cite{Christensen1996Jul, Grassberger2016Oct}, which has both mass and center of mass (CoM) conservation (analogous to dipole-moment conservation), but for which subdiffusive time-evolution equation, such as Eq. (1) of Ref. \cite{Han2023Apr}, does not hold.  
Our findings show that the dynamic properties of CoM conserving systems depend on microscopic details and need not always be subdiffusive.

The Oslo model, also known as the Oslo ``ricepile'' \cite{Christensen1996Jul}, is an example of threshold-activated systems, collectively called ``sandpiles'' \cite{bak_97}.
Not only does the model serve as a prototype for many-body systems with CoM conservation and broken time-reversal symmetry (thus no detailed balance), but it also constitutes a paradigm for absorbing-phase transition (APT) with multiple conservation laws.
Dynamical activities in these highly nonlinear  systems propagate in space and time via intermittent burst-like ``avalanches'', resulting in long-ranged spatio-temporal correlations.
Sandpiles have several variants, depending on whether mass variable being discrete or continuous, particle-transfer rules being deterministic or stochastic \cite{Dhar1990Apr, Dhar1999Feb, Manna_1991, Dickman2001Oct}, randomness in threshold mass \cite{Christensen1996Jul}, ``stickiness'' \cite{Mohanty2002Aug, Mohanty2007Oct} and various other local and nonlocal variations of microscopic dynamics \cite{Kadanoff1989Jun}.
The Oslo model is a stochastic variant of the BTW sandpile \cite{bak_97} in that the threshold mass can take random values; the model should be contrasted with the stochastic Manna sandpile \cite{Manna_1991}, which has a fixed threshold value (like the BTW sandpile) and the stochastic particle-transfer rule (unlike the Oslo model).
Because of the deterministic mass transfer, the CoM remains conserved in both the Oslo and BTW models.

Various conserved-mass transport processes, such as random-organization model \cite{Corte2008May}, chiral active matter \cite{LeiScAdv2019, KurodaJSTAT2023} and CoM conserving Manna sandpile \cite{Hexner2017Jan, Voituriez-NJP2019}, among others \cite{Ikeda2023Sep}, have been studied in the past to better understand the role of CoM conservation on their large-scale spatio-temporal properties in an out-of-equilibrium setting.
For concreteness, in this work we focus on the Oslo model, which, like other conserved-mass (``fixed-energy'') sandpiles, exhibits  an absorbing phase transition \cite{Lubeck2004Dec, Dickman2000Mar}, where the fraction $a(\rho)$ of {\it active} sites (those with mass greater than or equal to the threshold value) vanishes when density $\rho$ falls below a critical density $\rho_c$.
One of the most striking characteristics of such a transition is that, unlike in equilibrium, where fluctuations diverge for a continuous phase transition, mass fluctuations are suppressed further upon approaching criticality (from above) and, in the thermodynamic limit, eventually vanishes at the critical point. Such a phenomenon is known as {\it hyperuniformity} \cite{Torquato2003Oct, Torquato2016Aug, Torquato2018Jun, Torquato2021Nov} and has been receiving a lot of attention \cite{Pine2005Dec, Corte2008May, Jeanneret2014Mar, Hexner2015Mar, Weijs2015Sep, Grassberger2016Oct, Hexner2017Jan, Christensen-EPL2018}.
While time-series of avalanches in sandpiles have been extensively investigated in the past  \cite{Kertesz1990May, Manna1991Apr, Laurson_2005}, the time-dependent properties related to large-scale transport and relaxation have received much less attention \cite{Carlson_1990, Cunha_2009,  Yadav_2012, Garcia-Millan2018Jul}. Only recently, it has been shown that the relaxation of long-wavelength density perturbations in the celebrated Manna sandpile, which has only one conserved quantity (mass), is diffusive far from criticality, but exhibits transport instabilities near criticality \cite{Chatterjee2018Jun, Tapader2021Mar, Mukherjee2023Feb}. That is, the density-dependent transport coefficients - the self- and bulk-diffusion coefficients and the particle mobility - are all singular. Near criticality, while the mobility $\chi(\rho) \sim \Delta^{\beta}$ and the self-diffusion coefficient ${\cal D}_s(\rho) \sim \Delta^{\beta}$ vanish as a function of the relative density $\Delta=\rho-\rho_c$, the bulk-diffusion coefficient diverges $D(\rho) \sim \Delta^{-(1-\beta)}$, resulting in anomalous (superdiffusive) transport where the dynamic exponent $z = 2 - (1-\beta) / \nu_{\perp}$ was found to be related to the (static) order parameter exponent $\beta$ and correlation length exponent $\nu_{\perp}$ \cite{Chatterjee2018Jun, Tapader2021Mar}. 
A detailed study of the near-critical properties of the Oslo model was carried out in Ref.~\cite{Grassberger2016Oct} and the standard critical exponents were numerically found to be close to the rational values with $\beta \approx 5/21$, $\nu_{\perp} \approx 4/3$ and $z \approx 10/7$. However, the question of how multiple conservation laws affect the dynamic properties and the universality class \cite{LeDoussal2015Mar, Shapira2023Jun} is not fully understood yet.

In this paper, by using a microscopic approach, we provide a large-scale characterization of various static and dynamic properties of mass and current in the Oslo model far from as well as near criticality; we consider the system strictly above criticality. We begin with long-wavelength density relaxations, demonstrating that, despite having highly constrained microscopic dynamics due to center-of-mass conservation, the Oslo model exhibits diffusive relaxation away from criticality and super-diffusive relaxation near criticality.  
Next we study dynamic fluctuations by analytically calculating unequal-time (two-point) correlation functions, and associated power spectra, involving current and mass.
We find a mass-conservation principle, which connects (dynamic) current and (static) mass fluctuations and exactly determines the decay exponents of the near-critical dynamic correlation functions in terms of the standard static exponents. Away from criticality, we exactly calculate within the closure scheme the decay exponent.
Notably, the dynamic correlation functions are qualitatively different from that observed in diffusive systems with a single conservation law, and thus from that seen in the conserved Manna sandpiles.
 The findings are summarized below.

\begin{enumerate}[(I)]

\item \textit{Density relaxation.--} Far from criticality, by performing a diffusive scaling of space $X \rightarrow x=X/L$, time $t \rightarrow \tau=t/L^2$ and local density $\rho(X,t) = g(X/L^2,t/L^2)$ with $L$ being system size, we show that the space and time dependent coarse-grained density $g(x,\tau)$ satisfies a nonlinear diffusion equation $\partial_{\tau} g(x,\tau) = \partial_x [D(g) \partial_x g (x,\tau)]$, where $D(g) = d a(g) / dg$ and $a(g)$ are the density-dependent bulk-diffusion coefficient and the steady-state activity, respectively.
However, near criticality, the system becomes super-diffusive, with diverging $D \sim \Delta^{-(1-\beta)}$, thus implying relaxation time $\tau_r \sim L^2/D \sim L^z$ with the dynamic exponent $z = 2 - (1-\beta) / \nu_{\perp}$, a relation which is, quite remarkably, exactly satisfied by the critical exponents conjectured in Ref. \cite{Grassberger2016Oct}.

\item \textit{A mass-conservation principle and its consequences.--} We obtain a mass-conservation principle, which connects the time-integrated bond-current fluctuation $\abr{\intc_i^2(T)}$ and the static mass fluctuation $\abr{\Delta M_l^2}$ where $\Delta M_l = \br{M_l - \langle M_l \rangle}$, with $\intc_i(T)$ and $M_l$ being the cumulative current across $i$th bond in a time interval $[0,T]$ and mass in a subsystem of size $l$, respectively. Indeed, for any fixed $\rho > \rho_c$, the asymptotic values of current and mass fluctuations $\Sigma^2_Q(\rho)$  = $ \lim_{L \to \infty} \sbr{\lim_{T \to \infty} \abr{\intc_i^2(T)}}$ and $\Sigma^2_M(\rho) = \lim_{l \to \infty} \sbr{\lim_{L \to \infty} \abr{\Delta M_l^2}}$, respectively, converge to a finite value. By using a mass-conservation principle, they are related through an exact equality, 
  \begin{equation}
    \Sigma^2_Q(\rho) = \Sigma^2_M(\rho).
    \label{FR}
    \end{equation}
Consequently, the particle mobility $\chi(\rho) \equiv \lim_{L \to \infty} [\lim_{T \to \infty} L \abr{\intc_i^2(T, L)}/2T]$, which is defined by taking the infinite-time limit first and then the infinite-volume limit, vanishes identically for any density $\rho > \rho_c$, as opposed to the conserved Manna sandpile, or any other diffusive systems with a single conservation law, for which the mobility is finite (except at critical point).
 Interestingly, both the asymptotic current and mass fluctuations, $\Sigma^2_Q(\rho)$ and $\Sigma^2_M(\rho)$, respectively, are a nonmonotonic function of density; as a function of $\Delta = \rho - \rho_c$, they diverge as $ \Delta^2$ for $\Delta \gg \rho_c$ and as $ \Delta^{-\delta}$, with $\delta > 0$, for $\Delta \to 0^+$.

\item {\it Finite-time characteristics of time-integrated bond current.--} Temporal growth of time-integrated current fluctuation is qualitatively different from that in diffusive systems with a single conservation law. In the regime $1 \ll T \ll L^2$, we show that, far from criticality, the time-integrated bond-current fluctuation saturates as $\abr{\intc_i^2(T)} \simeq \Sigma^2_Q(\rho) - {\rm const.} T^{-1/2}$.
  However, near criticality, the current fluctuation grows as $\abr{\intc_i^2(T)} \sim T^\alpha$, but eventually saturates to a large $\Sigma^2_Q(\rho) \sim \Delta^{-\delta}$. By using eq.\eqref{FR}, we determine $\alpha = \delta / \nu_\perp z$ and $\delta = 2 \zeta/\nu_{\perp} = 2(1-1/\nu_\perp)/\nu_\perp$ in terms of the standard critical exponents - hyperuniformity, correlation-length and dynamic exponents $\zeta$, $\nu_{\perp}$ and $z$, respectively.

\item \textit{Statistics of instantaneous current.--} By decomposing instantaneous current $\mathcal{J}_i(t) \equiv \\d \mathcal{Q}_i(t)/dt$ as a sum of diffusive part $\jd$ and a fluctuating (or ``noise'') part $\jfl$, i.e., $\mathcal{J}_i(t) = \jd(t) + \jfl(t)$, we calculate respective dynamic correlation functions. We show that, far from criticality and for long times $t$, the current correlation function is long-ranged and decays as $\abr{\mathcal{J}_i(t) \mathcal{J}_i(0)} \simeq - {\rm const.} t^{-5/2}$, which is faster than that in the Manna sandpile, having a $t^{-3/2}$ power-law decay.
    However, the correlation function for the fluctuating current is delta correlated in time, i.e., $ \abr{\jfl_r(t) \jfl_0(0)} = \Gamma_r \delta(t)$, with space-integrated strength $\sum \Gamma_r = 0$, implying greatly suppressed current fluctuations, and thus resulting in the temporal and spatial hyperuniformity, in the system due to the CoM conservation.

\item \textit{Power spectrum of current and mass.--} Likewise, as $f \to 0$, we find that the power-spectrum of the instantaneous bond current $S_{\mathcal{J}}(f)  \sim f^{\psi_{\mathcal{J}}}$ with $\psi_{\mathcal{J}} = 1/2+\mu$, where $\mu = 1$ away from criticality and $\mu = 1/2 - \delta / \nu_\perp z$ near criticality; the decay is much faster compared to diffusive systems with a single conservation law, where $\mu=0$.
  Notably the above results imply that the current fluctuation far from criticality is much more suppressed than that near criticality.
  On the other hand, for small frequency, the power spectrum for subsystem mass $S_M(f) \sim f^{-\psi_M}$ grows with decreasing frequency, where the corresponding exponent is $\psi_{M} = 2 - \psi_{\mathcal{J}}$; similar to the current fluctuations, the mass fluctuations far from criticality are much more suppressed than that near criticality.

\item \textit{Static structure factor.--} We find that the static subsystem mass fluctuations are always hyperuniform, both far from and near criticality (strictly, from above), where the structure factor $S(q) \sim q^{\gamma}$ with $\gamma > 2$. Away from criticality, we analytically calculate, within our closure scheme, the structure factor and show that, for small wavenumber $q \to 0$, $S(q)$ is characterized by exponent $\gamma = 2$, thus exhibiting an extreme form of (``class I'') hyperuniformity \cite{Torquato2016Aug} in the active phase and thus providing a microscopic explanation for previous numerical observations in a CoM conserving system \cite{Hexner2017Jan}. Notably, the mass fluctuation is more suppressed far from criticality than that near criticality, consistent with the dynamic behavior of current and mass fluctuations [as described in points (II) - (V)].

\item \textit{Self-diffusion coefficient of tagged particles.--} Finally, we show that the density-dependent self-diffusion coefficient $\mathcal{D}_s$ exactly equals to the ratio of density-dependent activity to the density itself, i.e., $\mathcal{D}_s(\rhobar) = a(\rhobar) / \rhobar$. The result implies that, like conserved Manna sandpile \cite{Cunha_2009, Mukherjee2023Feb}, the behavior of the self-diffusion coefficient near criticality is exactly same as that of the order parameter - the activity. That is, like the activity, $\mathcal{D}_s(\Delta)$ decays as $\Delta^{\beta}$, where $\beta$ is the order-parameter exponent.

\end{enumerate}


We organize the paper as follows. In Sec. \ref{sec:model}, we define the Oslo model, and in Sec. \ref{sec-hydrodynamics}, we derive hydrodynamics of the model. Then, in Sec. \ref{sec-comparison-oslo}, we compare our hydrodynamic theory with simulations for both away and near critical density regimes. We describe the calculation of the current correlation in Sec. \ref{sec:integrated_current} and obtain general unequal-space and unequal-time integrated current fluctuations. Similarly, in Sec. \ref{sec:oslo_mass_ps}, we calculate the dynamic correlations involving mass and obtain a fluctuation relation. In Sec. \ref{sec:check_dynamic_correlation}, we compare our theoretical results for dynamic fluctuations with the simulations. In Secs. \ref{sec:oslo_ps} and \ref{sec:oslo_mass_ps}, we theoretically obtain and check the power spectra of current and mass. We calculate the self-diffusion coefficient of the Oslo model in Sec. \ref{sec:oslo_tagged_particle}, and in Sec. \ref{sec:oslo_structure_factor}, we calculate the static structure factor. Finally, we conclude with a summary in Sec. \ref{eq:oslo_summary}.


\section{Model}
\label{sec:model}

We consider the conserved-mass Oslo model - a prototype of conserved stochastic sandpiles \cite{Dickman2001Oct}, on a one-dimensional periodic lattice of size $L$. The continuous-time microscopic dynamical rule for a site $i \in [0$, $1$, $2$, $\ldots$, $L-1]$ is specified in terms of two local dynamical variables - the mass and the threshold mass for toppling. Mass (also called ``height'') or number of particles $m_i \ge 0$ at site $i$ takes integer values with total mass $N=\sum_{i=0}^{L-1} m_i$ conserved; the global density is denoted by $\rho = N/L$. The threshold value at site $i$ is denoted as $m_{c,i}$, with $m_{c,i} = 2$ or $3$,
reset to a new randomly chosen value after
a toppling occurs at the site. An {\it active} site $i$, with $m_i \geq m_{c,i}$, topples with unit rate by {\it deterministically} transferring two particles, one to its right-nearest neighbor and another to its left-nearest neighbor, followed by a random resetting of the threshold mass at site $i$.
The continuous-time update rules of local mass $m_i(t)$ at a site $i$ in an infinitesimal time interval $(t, t+dt)$ can be wriiten as
\begin{align}
    \label{eq:oslo_density_update_rules}
    m_i(t+dt) = 
    \begin{cases}
        \textbf{\textit{events}} &   \textbf{\textit{probabilities}} \\
        m_i(t)+1    &  \hata_{i+1} dt \\
        m_i(t)+1    & \hata_{i-1} dt \\
        m_i(t) - 2  & \hata_i dt \\
        m_i(t)  & \br{1-\Sigma dt},
    \end{cases}
\end{align}
where the sum of (exit) rates $\Sigma  = \br{\hata_{i+1} + \hata_{i-1} + \hata_i}$ and we denote $\hata_i$ as an indicator function,
\begin{align}
    \hat{a}_i = \begin{cases} 
  1 & \text{\textbf{\rm for} } m_i \geq m_{c,i}, \\
  0 & \text{\textbf{\rm otherwise}}. 
\end{cases}
\end{align}
The conserved Oslo model violates detailed balance (or, the microcopic time reversibility) in the bulk as the reverse transition corresponding to any allowed transition is forbidden. The system exhibits a continuous absorbing-phase transition (APT) below a critical density $\rho_c \approx 1.732594$ \cite{Grassberger2016Oct}; that is, for global density $\rho < \rho_c$, dynamical activities (topplings) cease to exist and the system settles into one of the many absorbing states, frozen in time and with the number of active sites being zero. On the other hand, for $\rho \geq \rho_c$, the system remains in an active state, where the dynamical activities go on for ever.
The absorbing phase transition is characterized through the steady-state density of active sites, or simply activity $a(\rho)$ - the ``order parameter'', which is computed as the steady-state average $a(\rho) = \sum_i \langle \hata_i \rangle / L$ and depends on the global density $\rho$.
Interestingly, the standard critical exponents for the Oslo model are known and, through large-scale simulations, they were found to be close to rational fractions, i.e., the order-parameter, correlation length and dynamic exponents are given by $\beta \approx 5/21$, $\nu_\perp \approx 4/3$, and $z \approx 10/7$, respectively \cite{Grassberger2016Oct}.

\section{Theory of Density relaxation}

\subsection{Hydrodynamics}
\label{sec-hydrodynamics}

In this section, we derive an exact hydrodynamic structure of the Oslo model. In contrast to the conserved Manna sandpile studied in Refs. \cite{Chatterjee2018Jun, Tapader2021Mar}, where particle transfer is stochastic, particle transfer rules in the Oslo model are deterministic with one particle moving to the right and the other to the left during a toppling event, thus conserving the CoM. While, for the Manna sandpile, the stochastic nature of particle transfer ensures diffusive relaxation (away from criticality), it is not evident that the density relaxation in the Oslo model will be diffusive as well. Indeed, for relaxation processes in the far-from-critical regime, we provide here a precise theoretical argument for why the relaxation process in the latter case is in fact diffusive. We also argue why a suitably coarse-grained density field must satisfy a nonlinear diffusion equation, quite similar to that arises in the case of the Manna sandpile \cite{Chatterjee2018Jun, Tapader2021Mar}.

We consider the Oslo model on a one dimensional periodic lattice having  $L$ sites. We specify the system through a local density $\rho_X(t) = \langle m_X(t)\rangle$ - the average of the particle number $m_X(t)$ at site $X$ and time $t$; or, equivalently, we can define a local excess density $\Delta(X, t) =(\rho_X(t) - \rho_c)$. 
By using the microscopic rules as in \eqref{eq:oslo_density_update_rules}, we can derive the following density-evolution equation, 
\begin{eqnarray}
\frac{\partial \rho_X(t)}{\partial t} &=& [a_{X-1}(t) - 2 a_X(t) + a_{X+1}(t)] 
\equiv \nabla^2 [a_X(t)], ~ ~~
\label{diffusion_discrete}
\end{eqnarray}
where $a_X(t)$ is the space and time dependent average local activity; here we denote $\nabla^2$ as the discrete Laplacian. It is worth noting that the density-evolution equation for the Oslo model is exactly the same in terms of appropriately defined corresponding local activities  \cite{Chatterjee2018Jun}.
Now we immediately find that the above equation for locally conserved density field can be expressed as a (discrete) continuity  equation ${\partial \rho_X(t)}/{\partial t} = J(X, t) - J(X+1,t),$  where we define a local current 
\be
J(X,t) = a_{X-1}(t) - a_X(t) \equiv - \nabla a_X(t);
\label{grad}
\ee 
note that here we have  the local current manifestly expressed as a discrete gradient of activity $a_X(t)$ and is readily recognized as diffusive current, which satisfies the Fick's law.
Importantly, we now invoke a local-equilibrium-like property of the inhomogeneous state so that the (``locally steady'') activity at long times can be essentially determined by the ``local steady-state'' activity,  which depends on the local coarse-grained density - the slow variable in the system. In other words, we write 
$a_X(t) = \langle \hat a_X \rangle_{\rho(X, t)}^{st} = a[\rho_X(t)],$ 
where $\langle . \rangle_{\rho_X(t)}^{st}$ represents the steady-state average conditioned on the fact that the local density is $\rho_X(t)$. 
As a result, the density evolution can be written in terms of a {\it nonlinear} diffusion equation,
\be
\frac{\partial \rho_X(t)}{\partial t} \simeq \frac{\partial^2 a[\rho_X(t)]}{\partial X^2}.
\label{non-lin-diff}
\ee
In the above equation, $a(\rho)$ is the activity calculated, in the steady state, as a function of density $\rho$.
Furthermore, the density-dependent bulk-diffusion coefficient can now be written as the derivative of (density-dependent) steady-state activity w.r.t. density, i.e.,  
\be 
D(\rho)=\frac{d a(\rho)}{d\rho}.
\label{D-rho}
\ee
The diffusive structure of equation \eqref{non-lin-diff} allows us to write the space and time dependent density in the scaling form,
\be 
\rho_X(t) = g \left( \frac{X}{L}, \frac{t}{L^2} \right),
\label{diff-scaling}
\ee
which, as one can immediately show, satisfies a nonlinear diffusion equation for coarse-grained density field $g(x,\tau)$,
\begin{equation}
\frac{\partial g(x,\tau)}{\partial \tau} = \frac{\partial^{2} a(g)}{\partial x^{2}} 
\label{diffusion_continuum}.
\end{equation}
Note that, in the limit of  $L$ large, the rescaled space $x$ and time $\tau$ in the above equation can be considered a continuous variable and the local coarse-grained activity $a(g)$ is a nonlinear function of $g$. In other words, the local-equilibrium hypothesis used to derive Eq. \eqref{diffusion_continuum} translates into the assertion that the local activity is indeed slave to local density.
Furthermore, on a large (coarse-grained) spatial scale, we can represent the initial density as a function of scaled position $x=X/L$, i.e., $\rho(X, t=0) \equiv \rho_{in}(X) = g_{in} \left( {X}/{L}\right).$ Clearly, the above equation must be solved given the initial condition $g(x,\tau=0) = g_{in}(x)$ and has a unique solution. Now, with a given density field, the CoM is also fixed locally (as well as globally) and therefore does not constitute an independent time-evolution equation for the other locally conserved quantity, i.e., the local CoM field.

It is quite expected that, away from criticality, correlation length $\xi$ in the system should be finite and hence the bulk-diffusion coefficient, as we find in this study, is nonzero and finite. In such scenario, the density perturbations, characterized by wave numbers $k \rightarrow 0$ being small, relax over a time scale $1/ k^2 D(\rho)$; equivalently, for a system with size $L$, we have the relaxation time $\tau_r \sim L^2/D(\rho)$. 
 The above mentioned argument for diffusive scaling, on the other hand, does not work in the near-critical regime. In fact, for near-critical density $\rho_X(t) \sim \rho_c^+$, the bulk-diffusion coefficient diverges as $D(\rho) \sim (\rho - \rho_c)^{\beta-1}$ because, in this scaling regime, the activity $a(\rho) \sim (\rho-\rho_c)^{\beta}$ has a form of a power-law with order-parameter exponent $\beta$ being model-dependent, but usually less than one \cite{Grassberger2016Oct, Dickman2001Oct}; in other words, the particle transport should be anomalous (superdiffusive) near criticality.

 Therefore we conclude that the hydrodynamic structure for density relaxation in the conserved-mass Oslo model is indeed quite similar to that in the case of the conserved Manna sandpile. This finding establishes one of the main results of this paper that, despite constrained dynamics \cite{Han2023Apr, Voituriez-NJP2019}, the density relaxation is {\it not always subdiffusive} in systems with CoM conservation. However, we must mention here that, due to the CoM conservation, the dynamic correlation functions for density and current fluctuations, on the contrary, are drastically different from that observed in the Manna sandpile \cite{Mukherjee2023Feb}.

 In the subsequent section, we obtain explicit theoretical solutions of Eq. \eqref{diffusion_continuum} (mostly obtained by a numerical integration scheme) for various initial conditions and compare them with direct Monte Carlo simulations. For the purpose of comparing the results for the Oslo model with that in the Manna sandpile, we consider similar parameter regimes (e.g., densities and system sizes, etc.) and the initial conditions for density field (e.g., step-like, wedge-like and Gaussian density profiles) as considered previously \cite{Tapader2021Mar}.

\subsection{Comparison: Theory and simulations}
\label{sec-comparison-oslo}

In this section, we explicitly obtain the solutions of the nonlinear diffusion equation (\ref{diffusion_continuum}), corresponding to various initial conditions $\rho(x, 0) \equiv g_{in}(x)$ in the domain $x \in [0,1]$; we consider throughout a periodic boundary condition $\rho(0)=\rho(1)$.
Starting from an initial density profile, we study the relaxation of density profiles on macroscopically large spatio-temporal scales. These initial density profiles, in most cases, are taken as a step, or box, density profile; though, in a few cases, wedge and Gaussian initial density profiles are also considered. 
In this way, we verify the nonlinear diffusion equation  (\ref{diffusion_continuum}), where we perform microscopic simulations in various density regimes (far from and near criticality) and compare the space and time dependent density with that obtained by theoretically solving equation (\ref{diffusion_continuum}).

To investigate the hydrodynamic theory, we numerically integrate  eq.\eqref{diffusion_continuum} by using the Euler method of numerical integration by discretizing space $x$ as $\delta x = 10^{-3}$ and time $\tau$ as $\delta\tau = 10^{-7}$, respectively. We integrate the  diffusion equation by using the explicit density dependence of the nonlinear function $a(\rho)$ - the steady-state activity. We generate the steady-state activity $a(\rho)$ vs. $\rho$ from direct microscopic simulations of the Oslo model and measure the activity in steps of density $\delta \rho = 10^{-2}$. In the numerical integration scheme, we  calculate $a(\rho)$,  by performing linear interpolation for densities having values smaller than the least possible value of density, i.e., for density values in the range $[\rho,\rho + \delta \rho]$.


\subsubsection{Relaxation of density profile far from criticality}
\label{sec-far-critical-oslo}

In this section, we study the density relaxation when the system is far from the criticality.  In simulation, to produce the initial density profile, we have generated numerous random initial configurations and done an ensemble average over them. Now, to observe the density profile at subsequent times, we evolve in time the initial density profile by obeying the microscopic dynamics up to that time and then averaging over initial (random) configurations and also over stochastic trajectories.

\begin{figure}[ht!]
\centering
\includegraphics[width=1.0\linewidth]{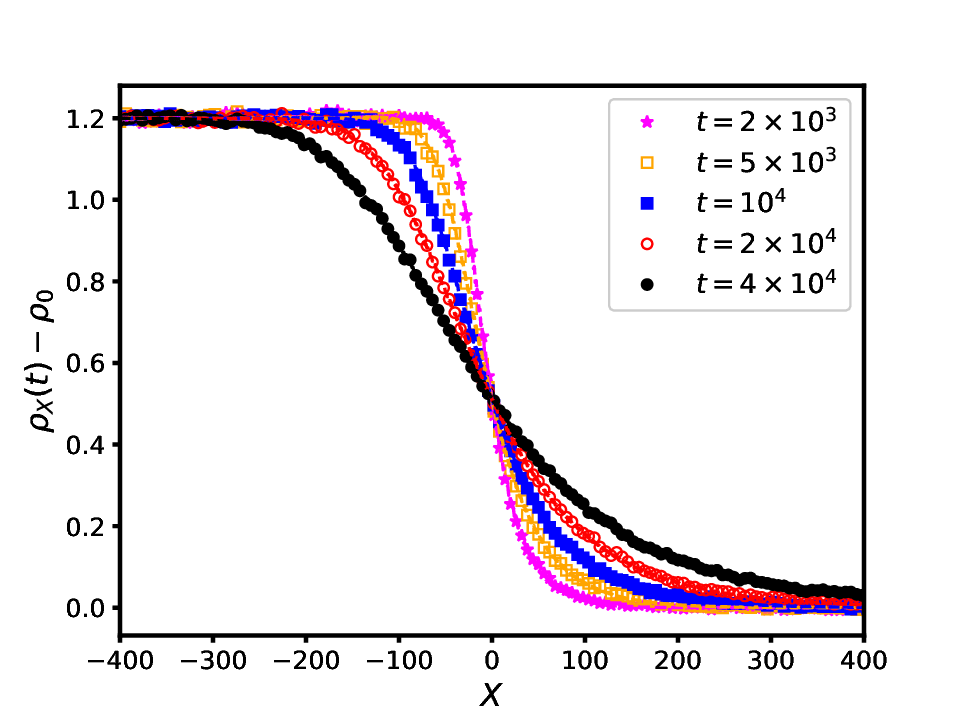}
\includegraphics[width=1.0\linewidth]{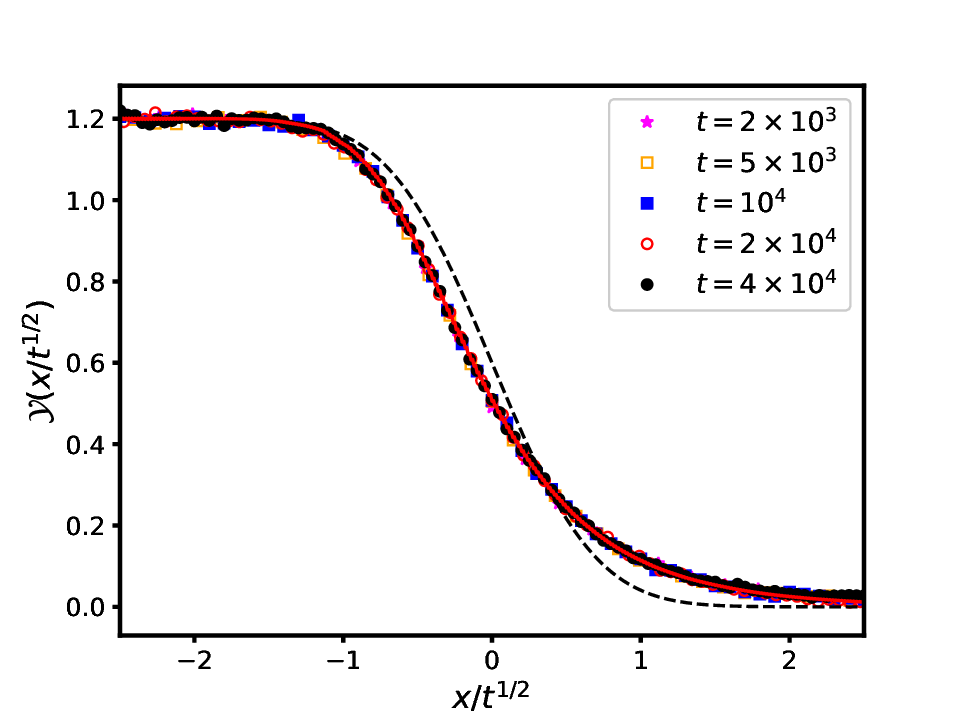}
\caption{ {\it Relaxation of a step-like density profile far from criticality on infinite domain.} \textit{Top panel:} Excess density over the uniform background density $\rho_{_X}(t)-\rho_{0}$ for a step initial profile \eqref{step-init-oslo} is plotted vs. position $X$. The background density is taken $\rho_{0}=1.8 > \rho_{c}$ and the height of the step on the left half of origin $\rho_{1}=1.2$. \textit{Bottom panel:} We plot the shifted density profile $\mathcal{Y}(z)$ vs. scaling variable $z=X/t^{1/2}$. Red-colored solid line represents the theoretical line obtained from the numerical solution of {\it nonlinear} Eq. (\ref{eq:scal_step}) and black-colored dashed line represents solution of the corresponding linear equation with a {\it constant} effective diffusivity (comparison purpose).  For both this panel, this step profile on the right half evolves and spreads over the domain, the profiles for different Monte Carlo times $t=2.0 \times 10^3$ (magenta-colored asterisks), $5.0 \times 10^3$ (orange-colored open squares), $10^4$ (blue-colored filled squares), $2.0 \times 10^4$ (red-colored open circles) and $4.0 \times 10^4$ (black-colored filled circles) are plotted. System size $L=2000$ and global density $\bar{\rho}=2.4$. Lines: hydrodynamic, points: simulation.}
\label{abv-rhoc-oslo}
\end{figure}

\textit{Relaxation of step initial profile.--} First we study the far-from-critical regime, where we consider the relaxation of a step  initial density profile, which spreads on an infinite domain having a density above the critical density. We take the step initial profile having height $\rho_{1}$ over a uniform density $\rho_{0} > \rho_{c}$ of the background. We are interested in characterizing how this initial step density perturbation relaxes around the origin $X=0$. To study this, we take an initial density profile having the form,
\begin{eqnarray}
\rho_{in}(X) =\left\lbrace
\begin{array}{ll} 
\rho_{1} + \rho_{0} & \mbox{for} ~ -\infty < X \leq  0, \cr
\rho_{0} & \mbox{otherwise.}           
\end{array}
\right.
\label{step-init-oslo}
\end{eqnarray} 
Here, $\rho_{1}$ is the height of the step, which is constructed over a uniform background density $\rho_{0}$ on the right half of the origin. We start with the following scaling ansatz where the (shifted) density profile,
\begin{align}
    \rho_{_X}(t)-\rho_0 = \mathcal{Y}\left(\frac{X}{\sqrt{t}}\right),
\end{align}
is written as a function of a scaling variable  $X/\sqrt{t}$.
Now, by substituting the above ansatz in Eq. (\ref{non-lin-diff}), one can show that the scaling function $\mathcal{Y}(z)$, with the scaling variable $z=X/t^{1/2}$, satisfies the following equation involving only a single variable $z$ [instead of two variables $X$ and $t$ in Eq. (\ref{non-lin-diff})],
\begin{align}\label{eq:scal_step}
    z\frac{d \mathcal{Y}(z)}{dz} = 2\frac{d}{dz}\left[D(\mathcal{Y})\frac{d\mathcal{Y}}{dz}\right],
\end{align}
which should be solved with the boundary conditions $\mathcal{Y}(-\infty) = \rho_1$ and $\mathcal{Y}(\infty)=0$. Of course, to find ${\cal Y}(z)$ analytically as an explicit function of $z$, we need to know the functional form of $D(\mathcal{Y})$, which is not known analytically.
However, Eq. \eqref{eq:scal_step} can be readily solved numerically and the scaling function ${\cal Y}(z)$ is plotted as a function of $z$ in Fig.\ref{abv-rhoc-oslo} (solid red line). Remarkably, our theory and simulations are in excellent agreement with each other, demonstrating that the far-from-critical relaxations are indeed governed by a {\it nonlinear} diffusion equation \eqref{diffusion_continuum}, where the bulk-diffusion coefficient is density-dependent.

In Fig. \ref{abv-rhoc-oslo},  the relaxation of step initial density profile \eqref{step-init-oslo} over an infinite domain is plotted and how this profile spread over the right half of the origin is studied. The global density of the system is taken $\bar{\rho}=2.4$ which is far away from criticality. So, it can be assumed that the transport of particles on the lattice will be diffusive resulting in a finite coefficient of diffusion of the system. We plot the shifted density profile $\rho(X,t)-\rho_{0}$ vs. position $X$ in top panel of Fig. \ref{abv-rhoc-oslo} for (Monte Carlo) times $t=2 \times 10^3$ (magenta colored asterisks), $5 \times 10^3$ (yellow-colored open squares), $10^4$ (blue-colored filled squares), $2 \times 10^4$ (red-colored open circles) and $4 \times 10^4$ (black-colored filled circles). In the bottom panel, we depict the scaling function of the shifted density profile as a function of the scaling variable $z$. The red solid line is obtained from the numerical solution of Eq. (\ref{eq:scal_step}), and it matches quite well with the simulation results, indicated by the data points in the plot.  We verify the density profiles obtained from microscopic simulation with the profiles from hydrodynamic equation \eqref{non-lin-diff}. We find that profiles from simulation (points) and hydrodynamic(lines) theory agree well.
Just to check the effect of nonlinearity, we also compare the actual solution ${\cal Y}(z)$ with the solution of  Eq. \eqref{eq:scal_step} with an effective (constant) bulk-diffusion coefficient $D_{eff} = D \br{\rho_{eff}}$ where $\rho_{eff} = \rho_0 + {\rho_1}/{2}$; one could see that the effective (constant $D$) solution $\mathcal{Y}_{eff}(z) = \rho_1 / 2 + \rho_1 \text{erf} \br{-z / {\sqrt{4 D_{eff}}}} / 2$, which is plotted in the same figure (dashed black line).

\subsubsection*{Verification of diffusive scaling limit}

The diffusive scaling limit used to arrive at eq. \eqref{diffusion_continuum} from eq. \eqref{non-lin-diff} is verified in this section. To check the scaling, we need to produce a scaled density profile $\rho(X=xL,t=\tau L^{2}) \equiv g(x,\tau)$, which is a function of scaled position $x=X/L^{2}$ for different system sizes and for different times, by keeping hydrodynamic time $\tau = t/L^{2}$ fixed. So, according to assertion that there exists a diffusive scaling limit, the density profiles for different system sizes and different times must collapse onto each other; moreover, the collapsed profile must also be described by the nonlinear diffusion equation \eqref{diffusion_continuum}.

Now to check the above mentioned scaling collapse in simulations, we take two different initial density profiles having a wedge and two steps (box). We take the following step-like initial density profile $ g_{in}(x) \equiv \rho \left(x, \tau=0 \right)$,
\begin{eqnarray}
g_{in}(x) =\left\lbrace
\begin{array}{ll} 
\rho_{1} + \rho_{0} & \mbox{for} ~ 0<x<x_{1}, \cr
\rho_{0} & \mbox{otherwise,}           
\end{array}
\right.
\label{step-diff-oslo}
\end{eqnarray}
where $\rho_{1}=7.0$ is the height of the step placed at the left quarter $x_{1}=1/4$ over a uniform background density $\rho_{0}=1.8$ where we keep the global density $\bar{\rho}=3.55$. The wedge-like initial profile is given by
\begin{eqnarray}
g_{in}(x) =\left\lbrace
\begin{array}{ll} 
\rho_0 + 2 \rho_1(x - x_1)/w & \mbox{for} ~ x_1 < x < x_2, \cr
\rho_0 + 2 \rho_1(x_3 - x)/w & \mbox{for} ~ x_2 < x < x_3, \cr
\rho_0 & \mbox{otherwise.}           
\end{array}
\right.
\label{wedge-diff-oslo}
\end{eqnarray}
We create the above wedge-like initial density profile by distributing $N_{1}=L(\bar{\rho} -\rho_{0})$ particles by keeping the center of the wedge at $x_2=1/2$. The width of the profile is taken to be $w=1/2$, from position $x_{1}=(1-w)/2$ to $x_{3}=(1+w)/2$. The height is taken to be $\rho_{1}=1.75$, placed over a background density $\rho_{0}=1.8$, chosen to be away from critical density $\rho_c$. In both of the cases, the global density is kept fixed at $\bar{\rho}=3.55$. For both the initial conditions, we check the diffusive scaling limit for system sizes $L=200,400,600$ and $1000$.

\begin{figure}[!ht]
\centering
\includegraphics[width=1.0\linewidth]{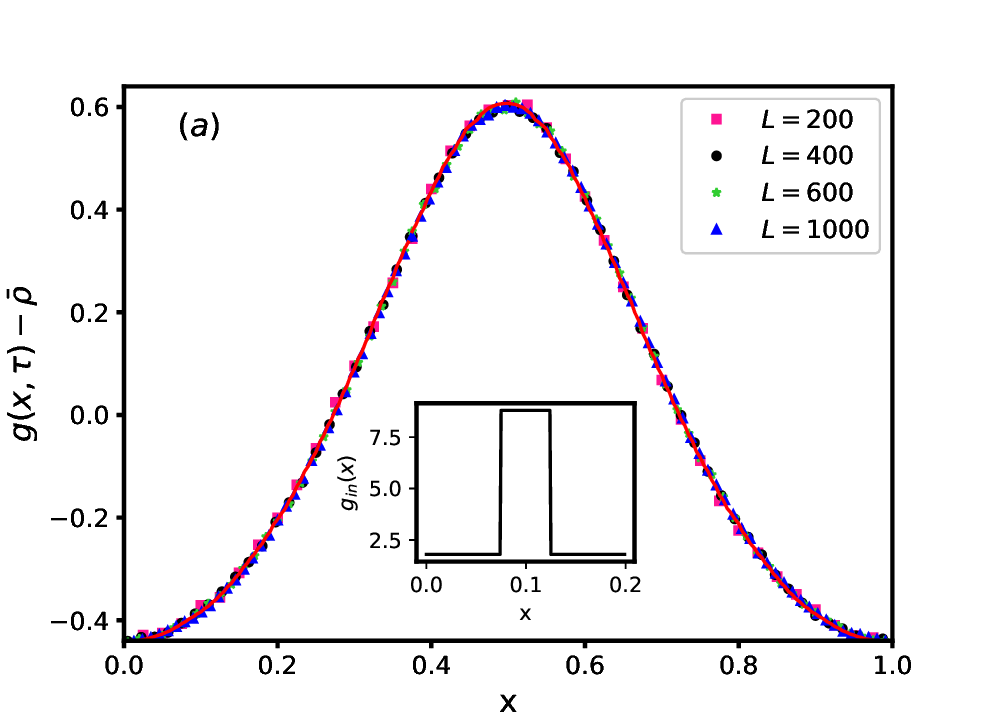}
\includegraphics[width=1.0\linewidth]{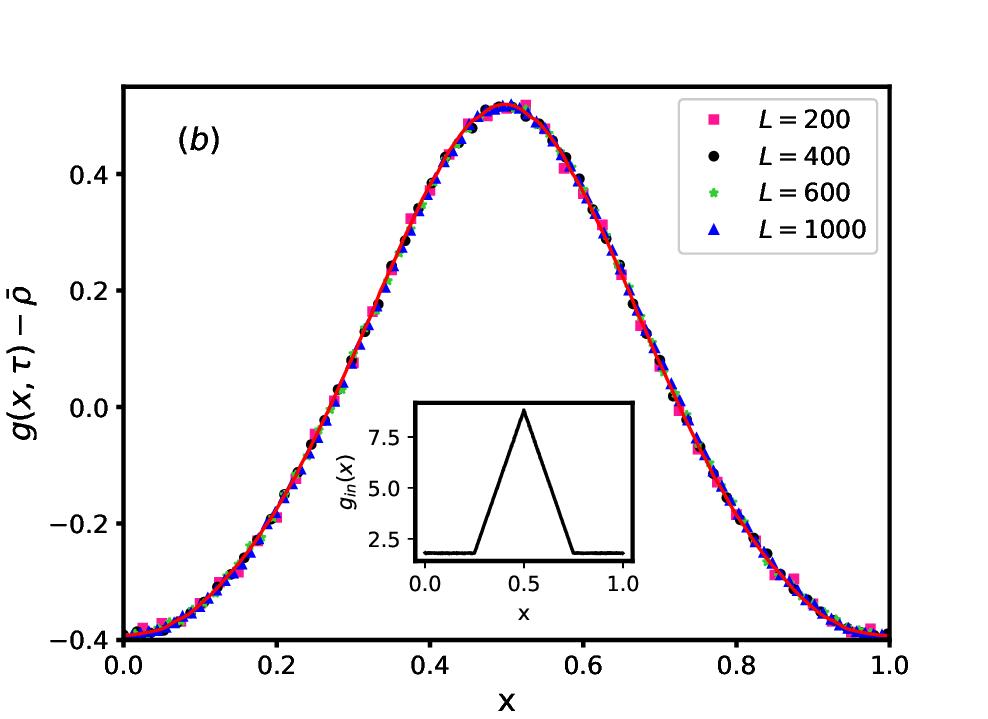}
\caption{{\it Verification of diffusive scaling limit in far-from-critical regime.} {Scaled density profile $g(x,\tau)-\bar \rho$ vs. rescaled position $x=X/L$ are plotted at times:   $t=2.0 \times 10^4$ for $L=200$ (pink-colored squares), $t=8.0 \times 10^4$ for $L=400$ (black-colored circles), $t=1.8 \times 10^5$ for $L=600$ (green-colored asterisks) and $t= 5.0 \times 10^5$ for $L=1000$ (blue-colored triangles), where the hydrodynamic time is kept fixed at $\tau=t/L^2=0.5$. Simulation points for different times are observed to collapse onto each other very well. In simulations, two sets of  $g_{in}(x)$ over a uniform background having density $\rho_{0}=1.8$ are considered: Step initial density perturbation [eq. \eqref{step-diff-oslo}] in panel (a) and wedge-like initial density perturbation [eq. \eqref{wedge-diff-oslo}] in panel (b). {\it Insets:} Initial scaled density $g_{in}(x)$ vs. scaled position $x=X/L$ is plotted. Lines represent theory, where numerical integration of eq. \eqref{diffusion_continuum} is performed, and points represent simulations.}}
\label{diffusive-oslo}
\end{figure}

 We present, in Fig. \ref{diffusive-oslo}, the scaled density profile $g(x,\tau) - \bar{\rho}$ over the global density as a function of scaled position $x = X/L$. We generate the profiles for a fixed hydrodynamic time $\tau = 0.5$, i.e., we have allowed the systems to evolve upto a Monte Carlo time $t = \tau L^2$ for respective system sizes. In simulations, with step initial profile presented in panel (a) and for wedge-like initial profile presented in panel (b), we evolve the density profiles up to times $t = 2 \times 10^4$ where $L = 200$ (pink-colored squares), $t = 8 \times 10^4$ where $L = 400$ (black-colored circles), $t = 1.8 \times 10^5$ where $L = 600$ (green-colored asterisks) and $t = 5 \times 10^5$ where $L = 1000$ (blue-colored triangles). We observe that all these density profiles collapse on to each other quite well. On the other hand, we obtain the scaled density function $g(x,\tau=0.5)$ by numerically integrating the hydrodynamic eq. \eqref{diffusion_continuum} upto the hydrodynamic time $\tau=0.5$ for both the initial conditions. From Fig. \ref{diffusive-oslo}, we find that the scaled density profile obtained from hydrodynamic theory and the collapsed density profiles obtained using simulations are in very good agreement.

\subsubsection{Near-critical density relaxation}
\label{sec-near-critical-oslo}

\begin{figure}[!ht]
\centering
\includegraphics[width=1.0\linewidth]{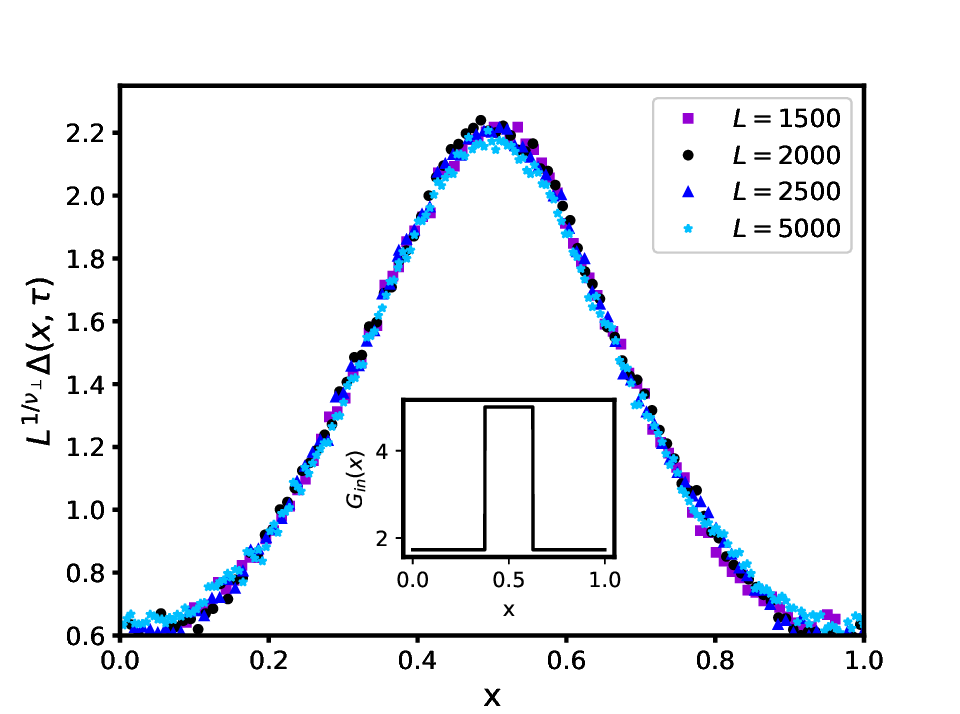}
\caption{{\it Verification of superdiffusive scaling near criticality.} We plot the scaled local excess density, denoted as $G(x, \tau) \equiv L^{1/\nu_{\perp}} \Delta(X=x L, t= \tau  L^z)$, against the scaled position $x$ for various times and system sizes: $t_1 = 10337$ for $L_1=1500$ (depicted by pink-colored squares), $t_2 = 15591$ for $L_2=2000$ (depicted by black-colored circles), $t_3 = 21445$ for $L_3=2500$ (depicted by blue-colored triangles), and $t_4=57725$ for $L=5000$ (depicted by sky-blue-colored asterisks). The hydrodynamic time is maintained constant at  $\tau=t/L^z \simeq 0.3$, with critical exponents set as $z = 10/7$, $\beta = 5/21$, and $\nu_{\perp} = 4/3$. Remarkably, the simulation points at different times and for various system sizes exhibit a very good scaling collapse. In the inset, we plot initial scaled excess density $G_{in}(x)$, defined as $G_{in} = L^{1/\nu_{\perp}} \Delta(X=x L, t=0)$, vs. scaled position $x=X/L$.}
\label{fig-finite_size_scaling-oslo}
\end{figure}

The near-critical regime corresponds to local excess density $\Delta(X,t) \sim L^{-1/\nu_\perp}$ very small; so the correlation length is large and it is of the order of the system size. As the correlation length is large, the transport is {\it not} diffusive any more. This is because the activity field $a(\rho) \sim \Delta^{\beta}$ has a form of a power law where we have exponent $\beta < 1$ and thus has a singularity at the critical point $\rho = \rho_{c}$. As a result, the bulk-diffusion coefficient, which is simply the derivative of the activity with respect to density, varies as $D(\Delta) \sim \Delta^{\beta 1}$, which diverges at the critical point. 
In the near-critical scaling regime, we use a finite-size scaling of the activity $A(\Delta, L) = L^{-\beta/\nu_{\perp}} {\cal A}(L^{1/\nu_{\perp}} \Delta)$, with ${\cal A}$ being a function of the scaling variable $L^{1/\nu_{\perp}} \Delta$  \cite{Dickman2001Oct, Mukherjee2023Feb}. Consequently, we obtain a time-evolution equation of the scaled excess density $G(x,\tau) = L^{1/\nu_{\perp}} \Delta(X, t)$, satisfying a non-linear diffusion equation  $\partial_{\tau} G(x,\tau) =  \partial_x^2 {\cal A}(G) $,
where space and time are scaled as following - $x = {X}/{L}$ and $\tau = {t}/{L^z}$, respectively; here $z$ is the dynamical exponent being determined via the two static exponents $\beta$ and $\nu_{\perp}$, where we use the following scaling relation,
\be
z=2 - \frac{(1-\beta)}{\nu_{\perp}}.
\label{SR}
\ee
The arguments given above can be verified in simulations.
To this end, we study rescaled excess density field $\mathcal{G}(x,\tau)$ starting from a step initial density profile $\mathcal{G}_{in}(x) \equiv \mathcal{G}(x,\tau = 0)$, which is given by
\begin{eqnarray}
G_{in}(x) =\left\lbrace
\begin{array}{ll} 
\rho_{1}  & \mbox{for} ~  x_1 \leq x \leq x_2, \cr
0 & \mbox{otherwise.}           
\end{array}
\right.
\label{step-super-diff_oslo}
\end{eqnarray}
We consider here the width of the initial profile $w = 1/4$  and height $\rho_{1} \simeq 5.0$ from $x_1 = 1/2 - w/2$ to $x_2 = 1/2 + w/2$. We generate the initial density profile over a uniform critical background density $\rho_{c}$.
Now, we verify the aforementioned `superdiffusive' scaling of the time-dependent density profiles. From this verification, we proceed to test the scaling relation as in \eqref{SR} as follows: We consider four systems with sizes $L_1 = 1500$, $L_2 = 2000$, $L_3 = 2500$, and $L_4 = 5000$. These systems evolve from an initial step profile given by \eqref{step-super-diff_oslo} up to times $t_1 = \tau L_1^{z}$, $t_2 = \tau L_2^{z}$, $t_3 = \tau L_3^{z}$, and $t_4 = \tau L_4^{z}$, with $\tau$ held fixed.
In this context, we determine the dynamic exponent $z$ by employing the scaling relation \eqref{SR}, by incorporating the previously conjectured static exponents $\beta \simeq 5/21$ and $\nu_{\perp} \simeq 4/3$ for the conserved Oslo model \cite{Grassberger2016Oct}. As part of our finite-size-scaling argument regarding the bulk-diffusion coefficient, we expect the density profiles, evolved up to the above mentioned  times, to exhibit a collapse onto each other.
In Fig. \ref{fig-finite_size_scaling-oslo}, we depict the scaled excess density $G(x,\tau)$, defined as $G \equiv L^{1/\nu_{\perp}} \Delta(X, t)$ and plot against the scaled position $x=X/L$ for four system sizes and times: $t_1 = 10337$ for $L_1=1500$ (represented by pink-colored squares), $t_2 = 15591$ for $L_2=2000$ (represented by black-colored circles), $t_3 = 21445$ for $L_3=2500$ (represented by blue-colored triangles), and $t_4=57725$ for $L=5000$ (represented by sky-blue-colored asterisks). Here, we set $\tau=t/L^z \simeq 0.3$ and our analysis involves averaging over $2 \times 10^5$ random initial configurations and trajectories. The results demonstrate a quite good collapse of the scaled shifted (excess) density profiles at the final above mentioned times.

\subsubsection{Density relaxation on critical background}

In this section, we study density relaxation on an infinite critical background having density $\rho_{c}$.
The excess density $\Delta$ over the critical background is not taken too far away from criticality but much larger than ${\cal O}(L^{-1/\nu_{\perp}})$ i.e., we have excess density of order ${\cal O}(L^{-1/\nu_{\perp}}) \ll \Delta \lsim 1$. So, the correlation length is still large $\xi \gg 1$ but $k \xi \ll 1$. In this regime, we exactly calculate, within our theory, the asymptotic scaling function for  density profiles, which have evolved upto a long time. Near criticality, the activity $a(\Delta)$ as a function of excess density $\Delta$ is a power law, i.e., we can write $a(\Delta) \simeq C \Delta^{\beta},$
where $C$ is the constant of proportionality and $\beta$ is the order parameter exponent. Now, we use the power-law scaling of activity in eq. \eqref{non-lin-diff}, and, in that case, the time-evolution equation for excess density can be written as following,
\begin{equation}
\frac{\partial \Delta(X, t)}{\partial t} = C\frac{\partial^2 [\Delta(X,t)]^{\beta}}{\partial X^2}.
\label{diffusion_equation_putting_activity-oslo}
\end{equation}
To study the density relaxation in this regime, we take initial condition having a form of delta function $\Delta(X, t=0) = N_1 \delta(X),$
with $N_{1}$ number of particles added at the midpoint,  to create the initial perturbation over the critical background. The reason for taking such initial condition is that here we are interested to study evolution of a localized initial density profile on an infinite critical background on a large space and time scales. In that case, it is interesting that, for the following boundary condition  $\Delta(x = \pm \infty, t) = 0$, the nonlinear diffusion equation \eqref{diffusion_equation_putting_activity-oslo} can exactly be solved.  We consider a scaling ansatz of the excess density as
\begin{equation}
\Delta(X, t) = \frac{1}{(C t)^{\omega}} {\cal G} \left[\frac{X}{(C t)^{\omega}}\right],
\label{scaling-fn-oslo}
\end{equation}
where the scaling function is ${\cal G}(y)$ satisfies the following differential equation,
\begin{equation}
\frac{d^2 {\cal G}^{\beta}}{dy^2} =  \omega \left[ {\cal G} + y \frac{d {\cal G}}{dy} \right],
\label{Gy-eq}
\end{equation}
where the growth exponent is given by $\omega={1}/(1+\beta)$; for the Oslo model, $\beta < 1$ and therefore $\omega > 1/2$.
By solving eq. \eqref{Gy-eq}, we exactly have
\begin{equation}
{\cal G}(y) = \frac{1}{\left[ g_0^{\beta - 1} + \frac{\omega(1 - \beta)}{2\beta} y^2 \right]^{1/(1 - \beta)}},
\label{gy-oslo}
\end{equation}
with the suitable choice of  boundary conditions, $ {\cal G}(y = 0) = g_{0}$ and $\left[ {d {\cal G}}/{dy}\right]_{y=0} = 0,$
where $g_{0}$ is the normalization constant, being proportional to the number of particles added initially in the system.

\begin{figure}[!ht]
\centering
\includegraphics[width=1.0\linewidth]{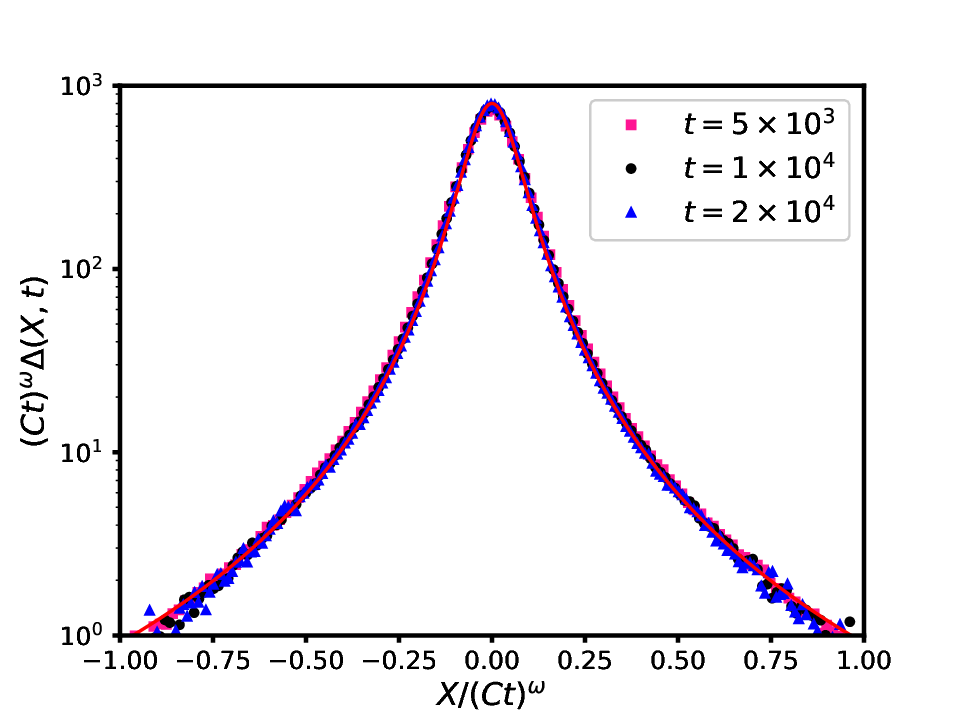}
\caption{{\it Density relaxation on critical background.} {We plot scaled excess density profile $(C t)^{\omega} \Delta(X,t)$, where $\Delta(X,t)=[\rho(X,t)-\rho_c]$,  against scaling variable $X/(C t)^{\omega}$ for times  $t=2 \times 10^4$ (blue-colored triangles), $10^4$ (black-colored circles) and $5 \times 10^3$ (pink-colored squares). The initial density profile is chosen to be a Gaussian one as in Eq. \eqref{gy-init-oslo}; we take $N_1=150$. Red line and points represent theory [Eq. \eqref{gy-oslo}] and simulations,  respectively.}}
\label{fig:gaussian_rlx}
\end{figure}

Next, we verify the above scaling solution in simulations. At initial time $t=0$, we consider a localized density profile, which has a Gaussian form,
\begin{equation}
\Delta(X, t=0) = N_1 \frac{1}{\sqrt{2\pi w^2}} e^{-X^2/2 w^2} ,
\label{gy-init-oslo}
\end{equation} 
where the width of the initial profile is taken $w = 10$ and in simulation $N_1 = 150$ number of particles are distributed according to the Gaussian distribution function \eqref{gy-init-oslo} over an infinite critical background with density $\rho_c$.
In simulation, a system size $L=5000$ is taken large in comparison to the width $\delta$ of the initial density profile and we also let the initial profile to relax for large times $t \gg 1$ such that, in the large spatial and temporal scales, the initial localized profile reduces to a Dirac-delta function. We compare simulations and the analytical solution as given in Fig. \ref{fig:gaussian_rlx} by plotting the scaled excess density ${\cal G}(y)$ as a function of $y$ for times $t =5 \times 10^3$ (pink-colored squares), $10^4$ (black-colored circles) and $2 \times 10^4$ (blue-colored triangles); the analytical form of the scaling function ${\cal G}(y)$ as in eq. \eqref{gy-oslo} is presented as red-colored line. We see from the plot that theory and simulation results are in excellent agreement.
This particular anomalous spreading in the Oslo model as encoded in eq. \eqref{gy-oslo}, with growth exponent $\omega > 1/2$, should be contrasted with the subdifusive scaling found in a CoM conserving system with broken time-reversal symmetry \cite{Voituriez-NJP2019}.

\section{Theory of Dynamic fluctuation}
\label{sec:fluctuation_properties}

\subsection{Current fluctuation}

\subsubsection{Definitions and notations}

Due to the conserved dynamics of the Oslo model, the microscopic time-evolution equation of local density, as provided in Eq. \eqref{diffusion_discrete}, can be expressed in a form of a microscopic continuity equation. This equation involves the local difference of the microscopic instantaneous particle current operator, denoted as $\mathcal{J}_i(t)$, across the bond $(i,i+1)$ in the temporal interval $(t,t+dt)$ and can be written as
\begin{align}
    \label{eq:oslo_micro_continuity_eqn}
    \frac{d \rho_i(t)}{dt}  = \abr{\mathcal{J}_{i-1}(t) - \mathcal{J}_{i}(t)},
\end{align}
where we write the average of instantaneous current as 
\begin{align}
    \label{eq:oslo_avg_Ji(t)}
    \abr{\mathcal{J}_i(t)} = \abr{\hata_i(t) - \hata_{i+1}(t)}.
\end{align}
We decompose the instantaneous current as a sum of diffusive current $\jd_i(t)$ and fluctuating current $\jfl_i(t)$,
\begin{align}
    \label{eq:oslo_J=Jd+Jfl}
    \mathcal{J}_i(t) = \jd_i(t) + \jfl_i(t),
\end{align}
where we identify the stochastic variable $\jd_i(t)$ as
\begin{align}
    \label{eq:oslo_Jd_expression}
    \jd_i(t) = \hata_i(t) - \hata_{i+1}(t),
\end{align}
with the following statistical properties of $\jd_i(t)$ and $\jfl_i(t)$,
\begin{align}
    \label{eq:oslo_Jd_stat_prop}
    &\abr{\jd_i(t)} \equiv \abr{\mathcal{J}_i(t)} = \abr{\hata_i(t) - \hata_{i+1}(t)}, \\
    \label{eq:oslo_Jfl_stat_prop}
    &\abr{\jfl_i(t)} = 0,
\end{align}
so that Eq. \eqref{eq:oslo_J=Jd+Jfl} is consistent with Eq. \eqref{eq:oslo_micro_continuity_eqn}. In this work, our main goal is to calculate dynamic correlation function $\abr{\mathcal{J}_i(t) \mathcal{J}_j(\tp)}$ and the fluctuation properties of $\jfl_i(t)$. To this end, we define time-integrated bond current $\intc_i(t)$, which represents cumulative (time-integrated) particle current across  bond $(i,i+1)$ in a time interval $\sbr{0,t}$ and is related to instantaneous current $\mathcal{J}_i(t)$ as
\begin{align}
    \label{eq:Q-J_rel}
    \mathcal{J}_i(t) = \left. \frac{d}{d\tp} \intc_i(\tp) \right\rvert_{\tp=t}.
\end{align}
From the dynamic correlations of  time-integrated current $\abr{\intc_i(t) \intc_j(\tp)}$, we can find the dynamic correlation of instantaneous current as
\begin{align}
    \label{eq:oslo_QQ-JJ-corr_def}
    \abr{\mathcal{J}_i(t) \mathcal{J}_j(\tp)} = \frac{d}{dt} \frac{d}{d\tp} \abr{\intc_i(t) \intc_j(\tp)},
\end{align}
for any arbitrary values of $t$ and $\tp$.  We introduce the following notation for correlation function,
\begin{align}
    \label{eq:oslo_CrAB(t,tp)-def}
    C^{A B}_r(t,\tp)  = \abr{A_{r}(t) B_{i+r}(\tp)} - \abr{A_{r}(t)} \abr{B_{i+r}(\tp)},
\end{align}
whereas the stationary correlation functions are defined as $C^{A B}_r(t) = C^{A B}_r(t,0)$. Also, here we define the Fourier transform of the correlation functions $C_r^{AB}(t,t^\prime)$ in the spatial domain of $r$ as 
\begin{align}
  \label{eq:oslo_first fourier}
  \tilde{C}_q^{AB}(t,t^\prime) = \sum\limits_{r=0}^{L-1}
    C_r^{AB}(t,t^\prime) e^{\imgi q r},
\end{align}
where we have $q=2\pi k/L$ with $k=0$, $1$, $\ldots$, $L-1$; then the inverse Fourier transform can be written as
\begin{align}
  \label{eq:oslo_inverse fourier}
  C_r^{AB}(t,t^\prime) = \frac{1}{L} \sum\limits_{q}
  \tilde{C}_q^{AB} (t,t^\prime) e^{- \imgi q r},
\end{align}
with $\imgi^2 = -1$.

\subsubsection{Correlation of integrated current and the truncation scheme}
\label{sec:integrated_current}

In this section, we study the unequal-space-time correlation of the integrated bond current $\intc$ starting from writing the evolution equations, which can be obtained from the following infinitesimal-time update rules for the quantity $\abr{\intc_i(t+dt) \intc_{i+r}(\tp)}$,
\begin{align}
    \label{eq:oslo_Qi(t+dt)Qj(tp)_update_rules}
    &\intc_i(t+dt) \intc_{i+r}(\tp) = \nonumber \\
    &\begin{cases}
        \textbf{\textit{events}} & \textbf{\textit{probabilities}} \\
        \br{\mathcal{Q}_i(t)+1} \intc_{i+r}(\tp) & \hata_i(t) dt \\
        \br{\mathcal{Q}_i(t)-1} \intc_{i+r}(\tp) & \hata_{i+1}(t) dt \\
        \mathcal{Q}_i(t) \intc_{i+r}(\tp) &  1 - \Sigma dt,
    \end{cases}
\end{align}
for $t > \tp$, where $\Sigma = \br{\hata_i(t) + \hata_{i+1}(t)}$. From the above update rules, we obtain the time-evolution equation for the two-point correlation function involving integrated bond current,
\begin{align}
    \label{eq:oslo_CrQQ(t,tp)-evl-eq-modified}
    \frac{d}{dt} C_r^{\intc\intc}(t,\tp) = \abr{\cbr{\hata_i(t) - \hata_{i+1}(t)} \intc_{i+r}(\tp)},
\end{align}
solving which, we obtain the exact expression for the unequal-time unequal-space correlation function of the integrated current, $C^{\intc \intc}_r(t,\tp)$, as
\begin{align}
    \label{eq:oslo_CrQQ(t,tp)-sol}
    C_r^{\intc\intc}(t,\tp) = \int_{\tp}^{t} d\tpp \abr{\jd_{i}(\tpp) \intc_{i+r}(\tp)} + C_r^{\intc\intc}(\tp,\tp),
\end{align}
where $\jd_{i}(\tpp)$ on the right-hand side of the above equation is given by Eq. \eqref{eq:oslo_Jd_expression}. Note that in Eqs. \eqref{eq:oslo_CrQQ(t,tp)-evl-eq-modified} and \eqref{eq:oslo_CrQQ(t,tp)-sol}, we omit all terms containing the average of the integrated current, $\langle\intc_i(t)\rangle$, since we are only concerned with the steady state where $\langle\intc_i(t)\rangle = 0$ due to the periodic boundary condition and the absence of any biasing force.

To further simplify the Eq. \eqref{eq:oslo_CrQQ(t,tp)-sol}, we need to calculate the unequal-time unequal-space correlation function for the diffusive current and the time-integrated current, which involves the dynamic correlation function of activity and integrated current. One could start by writing down the corresponding update rules for the function $\hata_i(t+dt) \intc_{i+r}(\tp)$ within an infinitesimal-time interval $(t,t+dt)$, and try to obtain the corresponding evolution equation, \newline $d\abr{\hata_i(t) \intc_{i+r}(\tp)}/dt$, which generates more higher-order correlation functions and thus we obtain an infinite hierarchy of correlation functions. 
Now we employ an approximate closure scheme along the lines of what was introduced in the context of the conserved Manna model \cite{Mukherjee2023Feb}. The closure scheme is incorporated by expressing the microscopic diffusive current across a bond in terms of the difference in local mass between two neighboring sites connected by that specific bond as
\begin{align}
    \label{eq:oslo_current_approximation}
    \jd_i(t) \simeq D(\rhobar) \br{m_i(t) - m_{i+1}(t)},
\end{align}
where $D(\rhobar) = a^\prime(\rhobar)$ is the bulk-diffusion coefficient as defined in Eq. \eqref{D-rho}.
In Eq. \eqref{eq:oslo_current_approximation}, we assume that any difference in local activity should follow the difference in local mass. This implies that the diffusive current is proportional to the local mass difference, with an additional assumption that fluctuations are quite small, allowing us to express the bulk diffusion coefficient obtained for the corresponding global density as the proportionality constant. The approximation in Eq. \eqref{eq:oslo_current_approximation} essentially implies that, involving any correlation function involving diffusive current and another dynamic quantity, we should substitute the diffusive current with Eq. \eqref{eq:oslo_current_approximation}; for example, consider the following correlation function,
\begin{align}
    \abr{\jd_i(t) A_{i+r}(\tp)} &= \abr{\cbr{\hata_i(t) - \hata_{i+1}(t)} A_{i+r} (\tp)}\nonumber\\ &\simeq D(\rhobar) \abr{\cbr{m_i(t) - m_{i+1}(t)} A_{i+r} (\tp)},
\end{align}
where $A$ could be any dynamic observable.
Now, by substituting Eq.\eqref{eq:oslo_current_approximation} in Eq.\eqref{eq:oslo_CrQQ(t,tp)-sol} we obtain,
\begin{align}
    \label{eq:oslo_CrQQ(t,tp)-sol-approx}
    C_r^{\intc\intc}(t,\tp) &\simeq a^\prime(\rhobar) \int_{\tp}^{t} d\tpp \abr{\cbr{m_i(\tpp) - m_{i+1}(\tpp)} \intc_{i+r}(\tp)}\nonumber \\& + C_r^{\intc\intc}(\tp,\tp),
\end{align}
which requires the correlation function $C_r^{m \intc}(t,\tp)$ instead of the correlation involving activity variable and integrated current. By using the following update rules of the function $m_i(t+dt) \intc_{i+r}(\tp)$ for $t > \tp$,
\begin{align}
    \label{eq:oslo_mi(t+dt)Qj(t)_update_rules}
    &m_i(t+dt) \intc_{i+r}(\tp) = \nonumber \\
    &\begin{cases}
        \textbf{\textit{events}} &   \textbf{\textit{probabilities}} \\
        \br{m_i(t)+1} \intc_{i+r}(\tp)    &   \br{\hata_{i+1} + \hata_{i-1}} dt \\
        \br{m_i(t) - 2} \intc_{i+r}(\tp)  & \hata_i dt \\
        m_i(t) \intc_{i+r}(\tp)  & \br{1-\Sigma dt},
    \end{cases}
\end{align}
where the probability of happening nothing is $\Sigma dt$, we obtain the corresponding evolution equation for $C_r^{m \intc}(t,\tp)$ as
\begin{align}
    \label{eq:oslo_mQ(t,tp)-evl-eqn}
    \frac{d}{dt} C^{m \intc}_r(t,\tp) &= \Delta_{i,k} \abr{a_k(t)\intc_{i+r}(\tp)} \nonumber \\
    &= \abr{\cbr{\jd_{i-1}(t) - \jd_i(t)} \intc_{i+r}(\tp)}.
\end{align}
where $\Delta_{i,k}=(\delta_{i+1,k} + \delta_{i-1,k} - 2\delta_{i,k})$ is the discrete Laplacian.
We further simplify the above equation by approximating the diffusive current by Eq. \eqref{eq:oslo_current_approximation} and obtain the following equation,
\begin{align}
    \label{eq:oslo_mQ(t,tp)-evl-eqn-approx}
    \frac{d}{dt} C^{m \intc}_r(t,\tp) \simeq a^\prime({\rhobar}) \sum_{k} \Delta_{i,k} \abr{m_k(t)\intc_{i+r}(\tp)}.
\end{align}
We can write the solution of the above equation using the discrete Fourier transformation (defined in Eq.\eqref{eq:oslo_first fourier}) as
\begin{align}
    \label{eq:oslo_mQ(t,tp)-sol-fourier}
    \tilde{C}^{m\intc}_q(t,\tp) = \exp\sbr{-a^\prime(\rhobar) \lambda_q (t-\tp)}
    \tilde{C}^{m\intc}_q(\tp,\tp),
\end{align}
where $\tilde{C}^{m\intc}_q(t,\tp)$ is the Fourier transformed $C^{m\intc}_r(t,\tp)$ and $\lambda_q = 2[1-\cos{q}]$ are the eigenvalues of the discrete Laplacian. 
Although Eq. \eqref{eq:oslo_CrQQ(t,tp)-sol-approx} along with Eq. \eqref{eq:oslo_mQ(t,tp)-sol-fourier} completely describe the dynamical correlations of the integrated current, we need to calculate the equal-time correlation of current and mass to obtain the complete solution.

By using the dynamical update rules of $m_i(t+dt)$ $ \intc_{i+r}(t+dt)$, we readily write the infinitesimal time-evolution equation for the equal-time two-point spatial correlation of mass and current, denoted as $C^{m\intc}_r(t,t)$, as following:
\begin{align}
    \label{eq:oslo_mQ(t,t)-evl-eqn}
    \frac{d}{dt} C^{m\intc}_r(t,t) \simeq a^\prime(\rhobar) \sum_k\Delta_{i,k} \abr{m_k(t) \intc_{i+r}(t)} + f_r(t),
\end{align}
where $f_r(t)$ is the source of the corresponding correlation function and is given as
\begin{align}
    \label{eq:oslo_f(r)}
    f_{r}(t) &= C^{m\hata}_r(t,t) - C^{m\hata}_{r+1}(t,t) \nonumber \\
    &+ a \cbr{3\br{\delta_{0,r+1} - \delta_{0,r}} + \br{\delta_{0,r-1} - \delta_{0,r+2}}};
\end{align}
for details, see appendix~\eqref{sec:oslo_appendix_mQ(t,t)-update-rules}. 
The solution of Eq. \eqref{eq:oslo_mQ(t,t)-evl-eqn}, obtained through the Fourier transformation, is given by,
\begin{align}
    \label{eq:oslo_mQ(t,t)-fourier-sol}
    \tilde{C}^{m\intc}_q(t,t) = \int_0^t d\tp \exp\sbr{-a^\prime(\rhobar) \lambda_q (t-\tp)} \tilde{f}_q (\tp),
\end{align}
where we write the Fourier transform of the source term as
\begin{align}
    \label{eq:oslo_fq_form}
    \tilde{f}_q =  a e^{-\imgi 2q} \br{e^{\imgi q} - 1}^3 + \tilde{C}_q^{m\hata} \br{1 - e^{-\imgi q}} .
\end{align}
We further calculate the correlation function $C^{m\hata}_r(t,t)$ by using the steady-state condition of equal-time two-point spatial mass correlation function, $d C_r^{mm}(t,t) / dt = 0$, from which we obtain
\begin{align}
    \label{eq:oslo_m(t)a(t)-sol}
    C^{m\hata}_r(t,t) = a \delta_{0,r} - \frac{a}{2} \br{\delta_{0,r-1} - \delta_{0,r+1}};
\end{align}
for details see appendix~\eqref{sec:oslo_appendix_equal-time-mass-mass_correlation}. Correspondingly, the Fourier transform of $C^{m\hata}_r(t,t)$, $\tilde{C}_q^{m\hata}$ is given by
\begin{align}
    \label{eq:oslo_m(t)a(t)-Fourier}
    \tilde{C}^{m\hata}_q(t,t) = a\frac{\lambda_q}{2},
\end{align}
and finally putting Eq. \eqref{eq:oslo_m(t)a(t)-Fourier} in Eq. \eqref{eq:oslo_fq_form} we write
$\tilde{f}_q$ as
\begin{align}
    \label{eq:oslo_mQ-source_fourier}
    \tilde{f}_q = -a \frac{\lambda_q}{2} \br{1-e^{-\iu q}}.
\end{align}
By substituting, Eq.\eqref{eq:oslo_mQ-source_fourier} in Eq.\eqref{eq:oslo_mQ(t,t)-fourier-sol} and then putting Eq.\eqref{eq:oslo_mQ(t,t)-fourier-sol} in Eq.\eqref{eq:oslo_mQ(t,tp)-sol-fourier}, we finally obtain the Fourier transform of unequal-time mass and integrated current as
\begin{align}
    \label{eq:oslo_mQ(t,tp)-fourier-sol-final}
    \tilde{C}^{m\intc}_q(t,\tp) = -a(\rhobar) \int_0^{\tp} d\tpp e^{-a^\prime(\rhobar) \lambda_q (t-\tpp)} \frac{\lambda_q}{2} \br{1-e^{-\iu q}}.
\end{align}
To obtain the complete expression of Eq. \eqref{eq:oslo_CrQQ(t,tp)-sol-approx}, we still need to calculate the equal-time correlation of the integrated current, which is given on the right-hand side of the corresponding equation. On the other hand, the first integral on the right-hand side is explicitly computed by using the inverse Fourier transform of Eq. \eqref{eq:oslo_mQ(t,tp)-fourier-sol-final}.
It can be shown from the microscopic update rules of the function $\intc_i(t+dt) \intc_{i+r}(t+dt)$, the equal-time integrated current correlation satisfies the following evolution equation,
\begin{align}
    \label{eq:oslo_QQtt_evl_eqn}
    \frac{d}{dt} C^{\intc \intc}_r(t,t) =& \Gamma_{r}(t) + C_r^{\jd \intc}(t,t) + 
    C_{L-r}^{\jd \intc} (t,t), \nonumber \\
    \simeq& \Gamma_{r}(t) + a^\prime \cbr{C_r^{m \intc}(t,t)- 
    C_{r-1}^{m \intc}(t,t)} \nonumber \\  &+ a^\prime \cbr{C_{L-r}^{m \intc}(t,t) - 
    C_{L-r-1}^{m \intc}(t,t)},
\end{align}
where we denote $\Gamma_r$ as the strength of the fluctuating current, or $\abr{\jfl_i(t) \jfl_{i+r} (\tp)} = \Gamma_r \delta(t-\tp)$; for a detailed proof of this relation see Ref. \cite{Mukherjee2023Feb}. Using microscopic dynamics, as demonstrated in appendix \ref{sec:oslo_appendix_equal-time-current-current_correlation}, the expression for $\Gamma_{i,j}$, or, equivalently, $\Gamma_r$ with $r=j-i$, can be written as
\begin{align}
    \label{eq:oslo_gamma_r}
    \Gamma_{i,j} = u_{i,j}(t) - u_{i+1,j}(t)
\end{align}
where $u_{i,j}(t)$ is given by
\begin{align}
    \label{eq:def_u}
    u_{i,j}(t) = \hata_i(t) \br{\delta_{i,j} - \delta_{i-1,j}}.
\end{align}
In the steady state, we can write $\Gamma_r$ simply as
\begin{align}
    \label{eq:oslo_gamma_r_sdst}
        \Gamma_{r} = 2a \delta_{0,r} - a \delta_{0,r+1} - a \delta_{0,r-1},
\end{align}
which is derived in Eq. \eqref{eq:oslo_QQ(tt)_source_sdst}. 
We obtain the full solution of Eq. \eqref{eq:oslo_QQtt_evl_eqn}, which constitutes the second part of Eq. \eqref{eq:oslo_CrQQ(t,tp)-sol-approx}, by substituting the inverse Fourier transform of Eq. \eqref{eq:oslo_mQ(t,t)-fourier-sol} for equal-time mass and integrated current correlation. Finally, using the inverse Fourier transform of Eq.\eqref{eq:oslo_mQ(t,tp)-fourier-sol-final} in Eq.\eqref{eq:oslo_CrQQ(t,tp)-sol-approx} and the solution of Eq.\eqref{eq:oslo_QQtt_evl_eqn}, we get the unequal-time unequal-space correlation function of the integrated current for $t \geq \tp$ as
\begin{widetext}
    \begin{align}
    \label{eq:oslo_CrQQ(t,tp)-sol-full-expression}
    C_r^{\intc\intc}(t,\tp) = \int\limits_0^{\tp} d\tpp  \Gamma_{r}(\tpp) &- 
    \frac{ a^\prime(\rhobar) a(\rhobar)}{L}\sum\limits_{q}
   \int\limits_0^{\tp} d\tpp \int\limits_0^{\tpp} d\tppp 
  e^{-a^\prime(\rhobar) \lambda_q (\tpp-\tppp)}
  \frac{\lambda_q^2}{2} \sbr{2-\lambda_{qr}}\nonumber \\ &- \frac{ a^\prime(\rhobar) a(\rhobar)}{L} \sum\limits_{q}
   \int\limits_{t}^{\tp} d\tpp \int\limits_0^{\tp} d\tppp 
  e^{-a^\prime(\rhobar) \lambda_q (\tpp-\tppp)}
  \frac{\lambda_q^2}{2} e^{-\imgi q r}.
\end{align}
\end{widetext}
The above expression of the dynamic current correlation function is subtly different from that in the Manna sandpile studied in Ref.~\cite{Mukherjee2023Feb} and has significant ramifications for the decay exponents of unequal-time current correlations and particle mobility.
To precisely understand the physical consequences of the above formula, and the corresponding fluctuating hydrodynamics of the Oslo model, we now study the detailed statistical properties of the space- and time-integrated current in the system.

\subsubsection{Fluctuating current and its relation to total current}
\label{sec:oslo_jfl_property}

Now we study the dynamic (two-point) correlation function for the fluctuating part $\jfl_i(t)$ of the instantaneous bond current, as defined in Eq.\eqref{eq:oslo_J=Jd+Jfl}, and we have
\begin{align}
    \label{eq:oslo fl current corr sol}
    C^{\mathcal{J}^{\br{fl}} \mathcal{J}^{\br{fl}}}_r(t,t^\prime=0)
    \equiv C^{\mathcal{J}^{\br{fl}} \mathcal{J}^{\br{fl}}}_r(t) =
    \delta(t) \Gamma_r(\rhobar),
\end{align}
where $\Gamma_r$ is given in Eq. \eqref{eq:oslo_gamma_r}. The above equation implies that the variance of the fluctuating current across a single bond is obtained by setting $r=0$ in Eq. \eqref{eq:oslo_gamma_r} as given below:
\begin{align}
    \label{eq:oslo_gamma_rule}
    \int_{-\infty}^{\infty}  C^{\mathcal{J}^{\br{fl}} \mathcal{J}^{\br{fl}}}_0(t)dt = 
    \Gamma_0 = 2a(\rhobar).
\end{align}
However, the variance of the total integrated fluctuating current is identically \textit{zero},  
\begin{align}
    \label{eq:oslo_gamma_sum_rule}
    \sum_r \int_{-\infty}^{\infty}  C^{\mathcal{J}^{\br{fl}} \mathcal{J}^{\br{fl}}}_r(t)dt = 
     \sum_r \Gamma_r = 0.
\end{align}
In a diffusive system, the variance $\abr{\bar{Q}^2(L,T)}$ $-$ $\abr{\bar{Q}(L,T)}^2$ of the space-time-integrated particle current $\bar{Q}(L,T) = \sum_{i=0}^{L-1} \mathcal{Q}_i(T)$ in a system of size $L$ and up to time $T$ in steady state is directly related to the density-dependent particle mobility, or, equivalently conductivity in a charged system, $\chi(\rho) = \lim_{L \rightarrow \infty}$ $\abr{\bar{Q}^2(L,T)} / 2LT $ \cite{Mukherjee2023Feb}. Alternatively, it can be shown that the actual current fluctuation is also related to the variance of fluctuating current as
\begin{align}
    \label{eq:oslo_total_current_fluctuation}
    \lim_{L \rightarrow \infty} \frac{\abr{\bar{Q}^2(L,T)}}{LT} = \sum_r \Gamma_r,
\end{align}
which, by using Eq.\eqref{eq:oslo_gamma_sum_rule}, immediately leads to the particle mobility to exactly vanish, i.e., $\chi = 0$ in the Oslo model. This result is consistent with our later finding that the current fluctuation $\abr{\intc_i^2(T)}$ saturates in the time domain $T \gg L^2$ [see Eq.\eqref{eq:oslo_Q2(T)_largetime}]. That is, for $T \gg L^2$, the time-derivative of the bond current, which also equals to $2\chi / L$ with $\chi$ being the mobility [see Eq.\eqref{eq:general_mobility_def}], is identically zero.

\begin{figure}[!ht]
    \centering
    \includegraphics[width=1.0\linewidth]{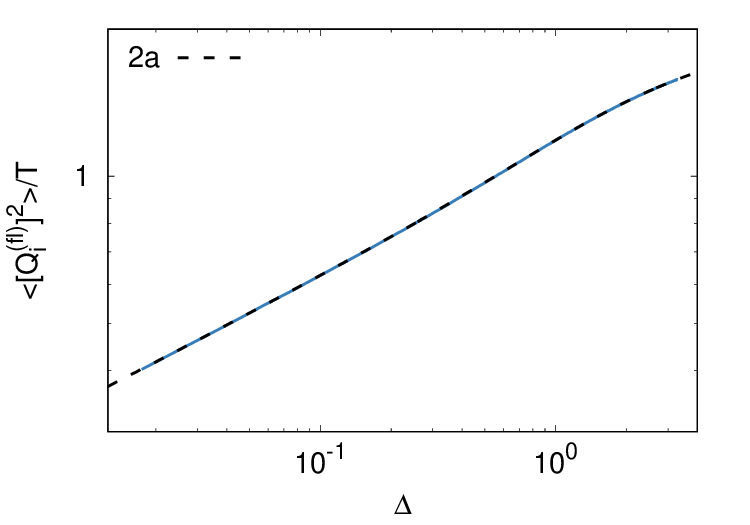}
    \caption{Intensive fluctuation of the excess bond current is plotted vs. relative density $\Delta = \rhobar - \rho_c$ ($\rho_c \approx 1.732$) for a system size of $L=1000$ and $T=100$, represented by the solid blue line. Additionally, we plot twice the activity $2a$ as a function of $\Delta$ in the same plot, represented by the dotted black line, which exhibits an excellent match with the intensive excess current fluctuation as described by Eq. \eqref{eq:oslo subsystem fl current variance}.}
    \label{fig:oslo_jfl_fluc_plot}
\end{figure}

To substantiate Eq.\eqref{eq:oslo_gamma_rule} in simulations, we define the following quantity - the time-integrated fluctuating bond current up to time $T$, 
\begin{align}
  \label{eq:oslo fl current numerical verification}
  \mathcal{Q}_i^{(fl)}(T) = \int\limits_0^T dt \mathcal{J}_{i}^{(fl)}(t).
\end{align}
Then, by using Eq. \eqref{eq:oslo_gamma_r}, we obtain a fluctuation  relation, which readily relates the scaled current fluctuation to the density-dependent activity,
\begin{align}
  \label{eq:oslo subsystem fl current variance}
  \frac{1}{T} \abr{\sbr{\mathcal{Q}_i^{(fl)}(T)}^2}
  =  2a(\rhobar).
\end{align}
In Fig. \ref{fig:oslo_jfl_fluc_plot}, we plotted the simulation data of $\abr{\sbr{\mathcal{Q}_i^{(fl)}(T)}^2} / T$ across a bond and up to time $T=100$ as a function of the relative density $\Delta = \rho - \rho_c$, shown as a solid blue line. We observe an excellent agreement with twice the activity, represented by the black dotted line, as given in Eq. \eqref{eq:oslo subsystem fl current variance}.

\subsubsection{Bond current fluctuation: Away from criticiality}
\label{sec:oslo_bond_current_fluctuation_away_from_criticality}

The steady-state fluctuation of the time-integrated bond current, $\abr{\intc_i^2(T)}$, can be obtained from Eq. \eqref{eq:oslo_CrQQ(t,tp)-sol-full-expression}, which by setting $r=0$ and $t=\tp=T$, results in,
\begin{align}
    \label{eq:oslo_Q2(T)}
    C_0^{\intc\intc}(T,T) = \abr{\intc^2_i(T,L,\rhobar)} = \frac{a(\rhobar)}{a^\prime(\rhobar)} \frac{1}{L} \sum_q \br{1 - e^{-a^\prime(\rhobar) \lambda_q T}}.
\end{align}
Now we perform the asymptotic analysis of the above equation. By rearranging the sum, we rewrite the above equation as
\begin{align}
    \label{eq:oslo_relative_Q2(T)}
     \abr{\intc_i^2(T)} = \frac{a}{a^\prime} - \frac{a}{a^\prime} \frac{1}{L} 
    \sum_q e^{-a^\prime \lambda_q T} + \mathcal{O}\br{\frac{1}{L}}, 
\end{align}
where we have used $\sum_q 1 / L \simeq 1$.
Now going to the continuum limit, we have
\begin{align}
    \label{eq:oslo_relative_Q2(T)_continuum}
     \abr{\intc_i^2(T)} \simeq \frac{a}{a^\prime} - 2\frac{a}{a^\prime} \int_{1/L}^{1/2} e^{-a^\prime\lambda(x) T}dx,
\end{align}
where $\lambda(x) \simeq 4\pi^2 x^2$. Furthermore, using the variable transformation $z = 4\pi^2 x^2 a^\prime T$, the above expression can be simplified as
\begin{align}
     \abr{\intc_i^2(T)} \simeq \frac{a}{a^\prime} - \frac{a T^{-1/2}}{2\pi {a^\prime}^{\frac{3}{2}}} \int_0^{\infty} dz \frac{e^{-z}}{\sqrt{z}} 
      = \frac{a}{a^\prime} - 
    \frac{a T^{-1/2}}{2\sqrt{\pi} {a^\prime}^{\frac{3}{2}}},
\end{align}
in the limit $T \gg 1$ being large. 
Note that, unlike the conserved Manna sandpile \cite{Mukherjee2023Feb} or the symmetric simple exclusion processes \cite{Sadhu2016Nov}, where there is only a single conserved quantity, the bond-current fluctuation $\abr{\intc^2_i(T,L,\rhobar)}$ in the Oslo model, for any fixed $\rhobar$ and in the limit of large time $T \gg L^2 \gg 1$, saturates to a finite value $\Sigma^2_Q(\rhobar)$. That is, we have the existence of the following limit,
\begin{align}
    \label{eq:oslo_Q2(T)_largetime}
     \lim_{L \rightarrow \infty} \left[ \lim_{T \rightarrow \infty} \abr{\intc^2_i(T,L,\rhobar)} \right] \equiv \Sigma^2_Q(\rhobar) = \frac{a}{a^\prime}.
\end{align}
The specific order of limits mentioned above means that we first take the infinite-time limit and then the infinite-system-size limit. For diffusive systems, the transport coefficient, which characterizes current fluctuation in the system, is the conductivity $\chi(\rhobar)$ (or equivalently, mobility) and is  defined as
\begin{equation}
\label{eq:general_mobility_def}
  \lim_{T \to \infty} \frac{ \abr{\intc^2_i(T,L,\rhobar)} }{T} \equiv 2 \frac{\chi(\rhobar)}{L}.
\end{equation}
It should be noted that, in the long-time limit ($T \gg L^2$), the vanishing temporal growth of fluctuations of the time-integrated bond current in the Oslo model renders the conductivity  $\chi(\rhobar) = 0$ - a direct consequence of the deterministic particle transfer during a toppling event.

\subsection{Mass fluctuation}
\label{sec:oslo_mass_ps}

The dynamic fluctuation properties of mass in a subsystem is another important quantity, that we study in this section.
To calculate the fluctuation of the mass of a subsystem of size $l$, $M_l(t) = \sum_{i=0}^{l=1} m_i(t)$, we begin with the steady-state correlation function of the unequal time and unequal space of the single-site mass for $t \geq 0$, $C_r^{mm}(t,0) = $ $\abr{m_i(t)m_{i+r}(0)} - \abr{m_i(t)}\abr{m_{i+r}(0)}$ $\equiv$ $C_r^{mm}(t)$. It is evident that, due to the microscopic dynamics that permit simultaneous mass changes in three sites, we can anticipate the presence of spatial correlations involving more than just individual sites in the steady state.
The corresponding evolution equation of $C_r^{mm}(t)$ can be written as
\begin{align}
    \label{eq:oslo_mimj(t,tp)_evl_eqn}
    \frac{d}{dt} C_r^{mm}(t) = \sum\limits_k \Delta_{i,k} \abr{\hata_k(t) m_{i+r}(0)},
\end{align}
using the following update rules for $t > \tp$,
\begin{align}
    \label{eq:oslo_crmm(t)_evolution_equation}
    &m_{i}(t+dt) m_{i+r}(\tp) =\nonumber \\ 
    &\begin{cases}
        \textbf{\textit{events}} & \textbf{\textit{probabilities}} \\
        \br{m_i(t)+1} m_{i+r}(\tp) & \hata_{i+1} dt \\
        \br{m_i(t)+1} m_{i+r}(\tp) & \hata_{i-1} dt \\
        \br{m_i(t)-2} m_{i+r}(\tp) & \hata_{i} dt \\
        m_{i}(t) m_{i+r}(\tp) & 1-\Sigma dt,
    \end{cases}
\end{align}
where $\Sigma dt = \br{\hata_{i+1} + \hata_{i-1} + \hata_{i}} dt$ is the probability of nothing happening.
Eq.\eqref{eq:oslo_mimj(t,tp)_evl_eqn} can be simplified further using the approximation scheme given in Eq.\eqref{eq:oslo_current_approximation} as
\begin{align}
  \label{eq:oslo_mimj(t,tp)_evl_eqn_approx}
  \frac{d}{dt} C_r^{mm}(t)
  \simeq {a^\prime} \sum\limits_k \Delta_{r,k} C^{mm}_{k}(t).
\end{align}
Solving the above equation in the Fourier modes, we obtain the following solution
\begin{align}
  \label{eq:oslo_Cqmm(t,t)_evl_eqn}
  \tilde{C}_q^{mm}(t) \simeq e^{-{a^\prime}(\rhobar) \lambda_q t}
  \tilde{C}_q^{mm}(0),
\end{align}
where the factor $\tilde{C}_q^{mm}(0)$ represents the Fourier transform of the (equal-time) two-point spatial correlation function. In the steady state, the corresponding equation involving this correlation function $C_r^{mm}(0)$, obtained by using the condition $d C_r^{mm}(t,t) / dt = 0$, can be written as
\begin{align}
  \label{eq:oslo_Cr(mm)_eqn}
  2{a^\prime} \sum\limits_k \Delta_{0,k} \abr{m_{k}m_{r}} + B_{r} = 0,
\end{align}
where the source term $B_r$ is given by
\begin{align}
    \label{eq:olso_Crmm_source_Br}
    B_r &= 6 a(\rhobar) \delta_{0,r} - 4a(\rhobar) \br{\delta_{0,r+1}+\delta_{0,r-1}}\nonumber \\ &+ 
    a(\rhobar) \br{\delta_{0,r+2}+\delta_{0,r-2}};
\end{align}
for details, see appendix \ref{sec:oslo_appendix_equal-time-mass-mass_correlation}. The corresponding generating function $\tilde{G}(z) = \sum_{z=0}^{\infty} C_r^{mm}(t,t) z^r$, to solve Eq.\eqref{eq:olso_Crmm_source_Br}, can be derived in a similar way to Eq. \eqref{eq:oslo_mass_mass_generating function}, by multiplying the above equation by $z^r$ and taking the sum over $r$ from $0$ to $\infty$, which finally leads to
\begin{align}
    \tilde{G}(z) = \frac{a}{\actp} - \frac{a}{2\actp}z.
\end{align}
The generating function leads to the steady-state spatial correlation function for mass,
\begin{align}
    \label{eq:olso mm correlation}
    C^{mm}_r(0) = \frac{a}{\actp} \delta_{0,r} - \frac{a}{2\actp} \br{\delta_{0,r-1} + \delta_{0,r+1}}.
\end{align}
The Fourier transform of the above correlation function can immediately be found as
\begin{align}
    \label{eq:chap2_Cqmm(t,t)}
    \tilde{C}^{mm}_q(t,t) = \frac{a}{2\actp} \lambda_q,
\end{align}
which we put in Eq.\eqref{eq:oslo_Cqmm(t,t)_evl_eqn} to obtain,
\begin{align}
  \label{eq:oslo_Cqmm(t,t)_sol}
  \tilde{C}_q^{mm}(t) \simeq e^{-\actp \lambda_q t} \frac{a}{2\actp} \lambda_q.
\end{align}
Finally, using the inverse Fourier transform of $\tilde{C}_q^{mm}(t)$, we get our desired correlation function $C_r^{mm}(t)$ as
\begin{equation}
  \label{eq:oslo_Crmm(t,t)_sol}
  C_r^{mm}(t) \simeq \frac{1}{L} \frac{a}{2\actp} \sum_{q}
  e^{-\iu q r} e^{-\actp \lambda_q t}
   \lambda_q.
\end{equation}
We can now compute the time-dependent correlation function for mass $M_l(t) = \sum_{r=0}^{l-1} m_r(t)$ in a subsystem of size $l$ (where we take $l < L$) by using the following standard identity,
\begin{align}
  \label{eq:oslo_subsystem_mass_corr_def}
  C^{M_l M_l}(t) = l C_0^{mm}(t) + \sum\limits_{r=1}^{l-1} (l-r)
  \br{C_r^{mm}(t) + C_{-r}^{mm}(t)}.
\end{align}
Then by substituting Eq.\eqref{eq:oslo_Crmm(t,t)_sol} into the above equation, we get the dynamic correlation function for mass of the subsystem as
\begin{equation}
  \label{eq:oslo_subsystem_mass_Cr(t)_sol}
  C^{M_l M_l}(t) \simeq \frac{1}{L} \frac{a}{2{a^\prime}} \sum_{q}
 e^{-{a^\prime} \lambda_q t} \lambda_{ql}.
\end{equation}
Using $\lambda_{ql}\simeq 2$ for all values of $q$ for $l \gg 1$, $L \gg 1$, and $l / L \ll 1$, and replacing the sum as an integral, followed by using the transformation $z=4 \pi^2 a' x^2 t$, we obtain the dynamical correlation function for the subsystem mass,
\begin{align}
  \label{cor_MM_t}
  C^{M_l M_l}(t) &\simeq \frac{2 a}{a'} \int_{1/L}^{1/2} e^{- a' \lambda(x)t} dx\nonumber\\
  &\simeq \frac{a}{2 \pi {a^\prime}^{3/2}} t^{-1/2} \int_0^{\infty} \frac{e^{-z}}{\sqrt{z}} dz
  =  \frac{a}{2 \sqrt{\pi} {a^\prime}^{3/2}} t^{-1/2}, ~~~~~~
\end{align}
which decays with time as $t^{-1/2}$.
Now, by taking the infinite time limit $t \to \infty$, the above equal-time correlation gives us the fluctuation of the subsystem mass,
\begin{equation}
\label{eq:mass_fluc_static}
  C^{M_l M_l}(0) = \abr{M_l^2} - \abr{M_l}^2 \equiv \Sigma^2_{M}(\rhobar, l),
\end{equation}
which, in the limit of subsystem size $l \rightarrow \infty$, converges to a finite value and can be written as a function of global density $\rhobar$ only,
\begin{align}
    \label{eq:oslo_subsystem_mass_fluc_rel}
    \Sigma^2_{M}(\rhobar) \equiv  \lim_{l \to \infty}  \Sigma^2_{M}(\rhobar, l)  = \frac{a}{{a^\prime}};
\end{align}
here we have already taken the thermodynamic limit $L \to \infty$ with $l/L \to 0$. In diffusive systems with a single conservation law (such as symmetric exclusion process or the Manna sandpile away from criticality), the subsystem mass fluctuation as in Eq.\eqref{eq:mass_fluc_static} grows as $l$, which is not the case here. As we discuss later in the context of static structure factor, this particular observation is related to an extreme form of hyperuniformity (class I).
Moreover, by comparing Eq.\eqref{eq:oslo_Q2(T)_largetime} and Eq.\eqref{eq:oslo_subsystem_mass_fluc_rel}, we immediately obtain the following identity,
\begin{align}
    \label{eq:oslo_new_fluctuation_relation}
    \Sigma^2_Q(\rhobar) = \Sigma^2_{M}(\rhobar),
\end{align}
which relates asymptotic (dynamic) time-integrated bond current fluctuation and the asymptotic (static) subsystem mass fluctuation.
We mention here that, while it captures the dynamic properties of the system remarkably well, the closure scheme, perhaps not surprisingly, is not strictly applicable in the near-critical regime since the spatial correlations in that case become long-ranged and are not captured in the calculations. However, the fluctuation relation as in eq. \eqref{eq:oslo_new_fluctuation_relation}, which is a direct consequence of a mass-conservation principle (discussed below), is found to be valid both near and far from criticality [see  Fig.\ref{fig:oslo_fluc_relation}]. Importantly, as argued later, eq. \eqref{eq:oslo_new_fluctuation_relation} can be used to determine the exponent governing the near-critical temporal growth of current in terms of the standard static exponents. 
Indeed the relation in eq. \eqref{eq:oslo_new_fluctuation_relation} could be understood physically from a mass-conservation principle as following. Let us consider a spatial domain $[0,l-1]$ of size $l$ and having mass $M_l(t)$ at time $t$. Now the time-integrated boundary current ${\cal Q}^{B}(t) = {\cal Q}_{-1}(t) - {\cal Q}_{l-1}(t)$ flown into the subsystem in a time interval $[0,t]$ is identically equal to the difference in the subsystem mass $\Delta M_l(t) = M_l(t) - M_l(0)$ at  two times $t=0$ and $t$. Therefore, the corresponding fluctuations are also equal, i.e., $\langle [{\cal Q}^{B}(t)]^2 \rangle = \langle [\Delta M_l(t)]^2 \rangle = \langle [M_l(0)]^2 \rangle + \langle [M_l(t)]^2 \rangle -2 \langle M_l(0) M_l(t) \rangle $; the equality is valid for any time $t$ and can be suitably generalized to higher dimensions. Now, by first taking the infinite-time limit and then the infinite-subsystem (and infinite-system) limit (i.e., $t \gg L^2$ and $L \gg l \gg 1$), and then assuming complete decorrelation between $M_l(t)$ and $M_l(0)$ [as shown in eq. \eqref{cor_MM_t}] as well as that between ${\cal Q}_{-1}(t)$ and ${\cal Q}_{l-1}(t)$, we recover the relation as given in eq. \eqref{eq:oslo_new_fluctuation_relation}. It should be noted here that the above scaling limit does not exist in the case of the conserved-mass Manna sandpile, and also other diffusive systems with a single conservation law, for which the time-integrated boundary current fluctuations $\langle {\cal Q}^2_l(t) \rangle$ or $\langle [{\cal Q}^{B}(t)]^2 \rangle$, and therefore $\langle [\Delta M_l(t)]^2 \rangle$, grow with time.

\subsection{Comparison: Theory and simulations}
\label{sec:check_dynamic_correlation}

\subsubsection{The mass-conservation principle and its consequences}
\label{sec:fluc_rel}

\begin{figure}[!ht]
	\centering
    \includegraphics[width=1.0\linewidth]{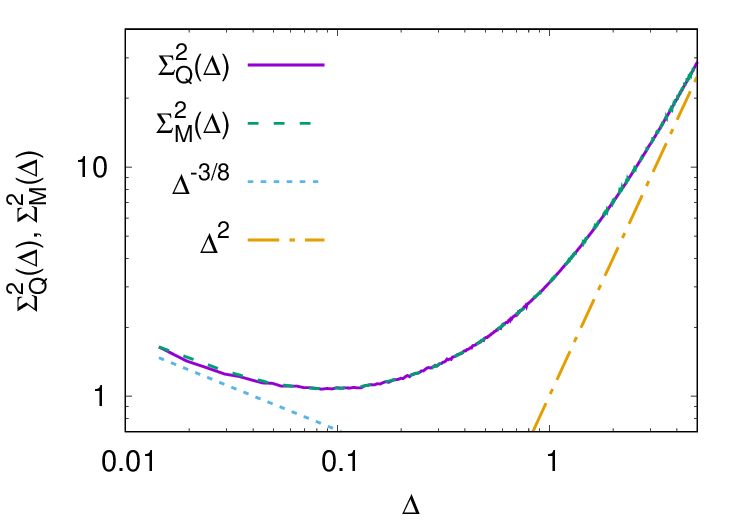}
    \caption{Comparison of bond current fluctuation $\Sigma^2_Q(\Delta)$ and subsystem mass fluctuation $\Sigma^2_{M}(\Delta)$ are plotted vs. relative density $\Delta = \rhobar-\rho_c$. Simulation data for current fluctuation is shown in a solid violet line for $L=1000$. Corresponding simulation data for $\Sigma^2_M(\Delta)$ is shown in a dashed green line for $L=1000$ and $l=500$, which shows excellent agreement with the bond current fluctuation, confirming our theoretical prediction of Eq. \eqref{eq:oslo_new_fluctuation_relation}. The dotted guiding line on the left, near critical density regime, indicates the $\Delta^{-3/8}$ density dependence of the fluctuations, whereas, the dot-dashed line on the right, indicates the $\Delta^2$ growth of the fluctuations away from criticality regime.
    }
     \label{fig:oslo_fluc_relation}
\end{figure}

To numerically verify the relation in Eq. \eqref{eq:oslo_new_fluctuation_relation}, we plot, in Fig. \ref{fig:oslo_fluc_relation}, the simulation data for $\Sigma^2_Q(\rhobar)$ for $L=1000$ and $T = L^2$ in a solid violet line and the corresponding simulation data for $\Sigma^2_{M}(\rhobar)$ in a dashed green line for $L=1000$ and $l=500$, both as a function of $\Delta = \rhobar-\rho_c$. The comparison between the two data sets substantiates the validity of the relation $\eqref{eq:oslo_new_fluctuation_relation}$ throughout the active phase of the model, far from and near criticality. From the same figure, we find that the density dependence of $\Sigma^2_{M}$ has two distinct regimes - one which is far from criticality and the other one near criticality. The qualitative behavior of mass fluctuation as a function of density is given by
\begin{align}
    \label{eq:oslo_subsystem_mass_fluc}
    \Sigma^2_{M}(\Delta) \sim 
    \begin{cases}
    		\Delta^2 & \text{for} \hspace{6pt} \Delta \gg \rho_c \\
    		\Delta^{-\delta} & \text{for} \hspace{6pt} \Delta \to 0^+
    \end{cases},
\end{align}
where the exponent $\delta$ has not yet been reported in the literature and is determined below in terms of the standard exponents.
\textit{Away from criticality}, we observe from Eq. \eqref{eq:oslo_subsystem_mass_fluc_rel} that the 
subsystem mass fluctuation does not depend on the size of the subsystem, i.e., the fluctuation is highly suppressed. By using Eq. \eqref{eq:olso mm correlation}), we see that such an extreme suppression of mass fluctuation is because the integrated density correlations vanish and the subsystem mass fluctuation only depends on the correlation among masses at the boundary sites.
Due to this particular behavior of $\Sigma^2_M(\Delta)$, the density fluctuation can be regarded as maximally hyperuniform, i.e., far from criticality, we observe \textit{class I} hyperuniformity \cite{Torquato2016Aug, Torquato2018Jun} in the Oslo model. We shall discuss this point later when we calculate the structure factor in this regime.
Furthermore, away form criticality, the activity is expected to behave as $a(\Delta) \sim 1 - \text{const.}/\Delta$, using which in Eq.\eqref{eq:oslo_subsystem_mass_fluc_rel}, we obtain  $\Sigma^2_M(\Delta) \sim \Delta^2$ for $\rhobar \gg \rho_c$. 
On the other hand, near-critical behavior of $\Sigma^2_M(\Delta)$ is obtained by using  Eq.\eqref{eq:oslo_new_fluctuation_relation} as following.
It is known that, as $\Delta \to 0$, the subsystem mass fluctuation becomes hyperuniform and scales with  system size $L$ as \cite{Grassberger2016Oct}
\begin{align}
	\label{eq:oslo_hyperuniformity}
	\Sigma^2_{M} \sim L^{\zeta} \sim L^{2(1-1/\nu_\perp)},
\end{align}
where $\zeta=2(1-1/\nu_\perp)$ and $\nu_\perp$ are the hyperuniformity and correlation length exponents, respectively. 
We obtain the exponent $\delta$ by using the finite size scaling relation $L \sim \Delta^{-\nu_\perp}$ in Eq.\eqref{eq:oslo_hyperuniformity} as
\begin{align}
    \label{eq:oslo_fluctuation_relation_exponent}
    \delta = \frac{\zeta}{\nu_{\perp}} = \frac{2(1-1/\nu_\perp)}{\nu_\perp} = \frac{3}{8},
\end{align}
where $\nu_\perp = 4/3$ \cite{Grassberger2016Oct}. Now, by using the relation Eq.\eqref{eq:oslo_new_fluctuation_relation}, we immediately obtain the following near-critical long-time behavior of bond-current fluctuation,
\begin{align}
    \label{eq:oslo_near_criticality_Q(Delta)}
    \Sigma^2_Q(\Delta) \sim \Delta^{-{3}/{8}}.
\end{align}

\begin{figure}[!ht]
    \centering
    \includegraphics[width=1.0\linewidth]{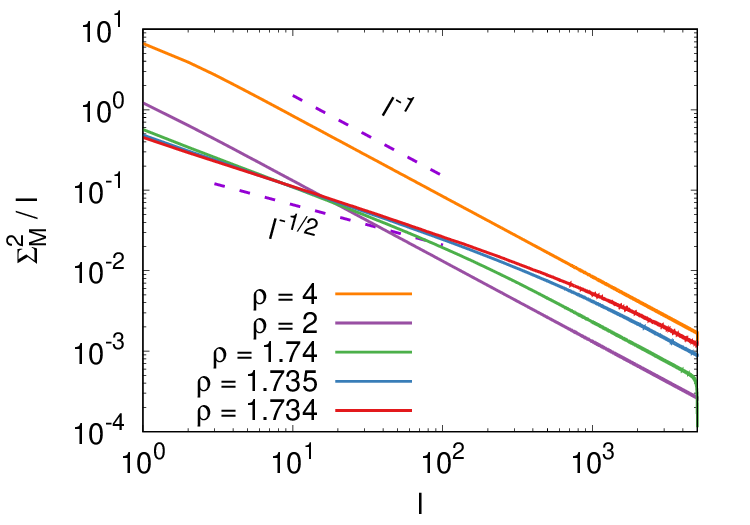}
    \caption{
    Scaled mass fluctuation, $\Sigma^2_{M}(\rho,l) / l$ vs. subsystem size $l$ is plotted. Simulation data for $\rhobar=4$ (orange line), $2$ (violet line), and $1.74$ (green line) are obtained for $L=5000$ and plotted in solid lines, whereas data for $\rhobar=1.735$ (blue line) and $1.734$ (red line) are plotted for $L=10000$. Away from criticality, $\Sigma^2_M(\rho,l) / l$ decays as $l^{-1}$ and indicates $\Sigma^2_M (\rho,l)$ does not depend on the subsystem size $l \gg 1$ (upper guiding line). Near criticality, $\Sigma^2_M(\rho,l) / l \sim l^{-1/2}$, implying $\Sigma^2_M(\rho,l) \sim l^{1/2}$ (lower guiding line) as found in \cite{Grassberger2016Oct}.}
    \label{fig:oslo_fluctuation_relation_check}
\end{figure}

We plot scaled fluctuation of subsystem mass $\Sigma^2_M / l$ as a function of the subsystem size $l$ in Fig. \ref{fig:oslo_fluctuation_relation_check}. Simulation data for $\rhobar=4$ (violet line), $2$ (green line) and $1.74$ (blue line) are shown for $L=5000$, while $\rhobar=1.735$ (orange line) and $1.734$ (red line) are plotted for $L=10000$. Away from the criticality, it indicates $l^{-1}$ decay of the scaled fluctuation or the maximal possible hyperuniformity or known as \textit{class I}  hyperuniformity \cite{Torquato2016Aug} in a \textit{one-dimensional} system, verifying our theoretical result obtained in Eq. \eqref{eq:oslo_subsystem_mass_fluc}. In contrast, near criticality the decay becomes much slower $l^{-\zeta}$, with $\zeta \simeq 0.5$, indicating hyperuniform fluctuations. The slight deviation from the value $\zeta=1/2$, as reported in \cite{Grassberger2016Oct}, is due to the finite-size effects.
The transition from an extreme form of (\textit{class I}) hyperuniformity with $\zeta = 1$ to hyperuniformity with $\zeta = 1/2$ \cite{Grassberger2016Oct} indicates an increase in near critical fluctuations not only in mass but also in current; this particular feature can be understood in the light of  Eq. \eqref{eq:oslo_new_fluctuation_relation}, which indicates a non-monotonic growth of fluctuation and a growing (eventually diverging at criticality) length scale in the system. This observation is consistent with our results on instantaneous current fluctuation and power spectrum discussed later in Sec. \ref{sec:oslo instantaneous_current_fluc} and \ref{sec:oslo_ps}.

\subsubsection{Bond current fluctuation: Away from criticality}

\begin{figure}[!ht]
    \centering
    \includegraphics[width=1.0\linewidth]{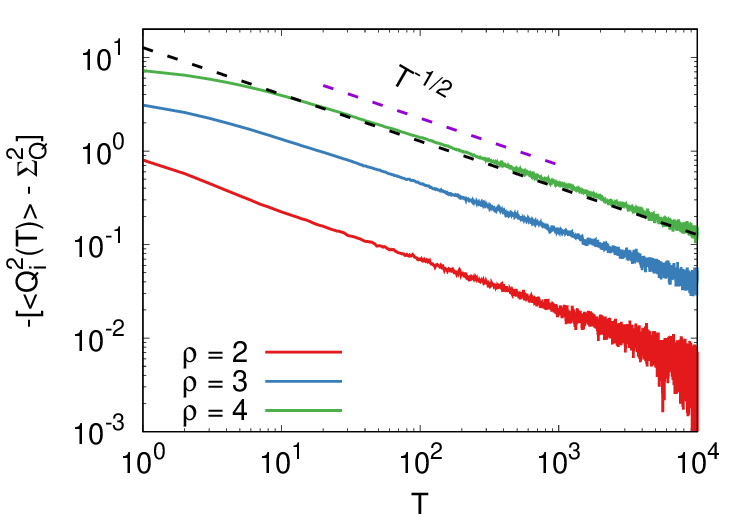}
    \caption{Relative integrated bond current fluctuation, $-\sbr{\abr{\intc_i^2(T)} - \Sigma^2_Q}$, is plotted for different densities as a function of time for system size $L$ $=$ $1000$. Simulation data are plotted in solid lines for densities $\rhobar=2$ (red line), $\rhobar=3$ (blue line), $\rhobar=4$ (green line) from bottom to top. We plotted the asymptotic-value of the fluctuation in  dotted-line for $\rhobar=4.0$ using Eq.\eqref{eq:oslo_Q2(T)_largetime}, which precisely capturers the $T^{-1/2}$ decay of the relative fluctuation of the current with time.
}
    \label{fig:oslo_away_criticality_Q2(T)}
\end{figure}

In Fig. \ref{fig:oslo_away_criticality_Q2(T)}, we plot the relative integrated current fluctuations, $-\sbr{\abr{\intc_i^2(T)} - \Sigma^2_Q}$ as a function of $T$. In the solid lines, from bottom to top, we plot the simulation data for the density values $\rhobar = 2$ (red), $3$ (blue), and $4$ (green), for system size $L=1000$. The corresponding asymptotic expression of $-\sbr{\abr{\intc_i^2(T)} - \Sigma^2_Q}$ for $\rhobar=4.0$ is plotted in black-dotted line using Eq.\eqref{eq:oslo_Q2(T)_largetime}, which shows an nice agreement with the theory and simulation as the time increases.
However, near criticality from simulation we observe that the current fluctuation changes dramatically, which we discuss in the next section.

\subsubsection{Bond current fluctuation: Near criticality}
\label{sec:oslo_bond_current_fluctuation_near_criticality}

\begin{figure}[!ht]
    \centering       
    \includegraphics[width=1.0\linewidth]{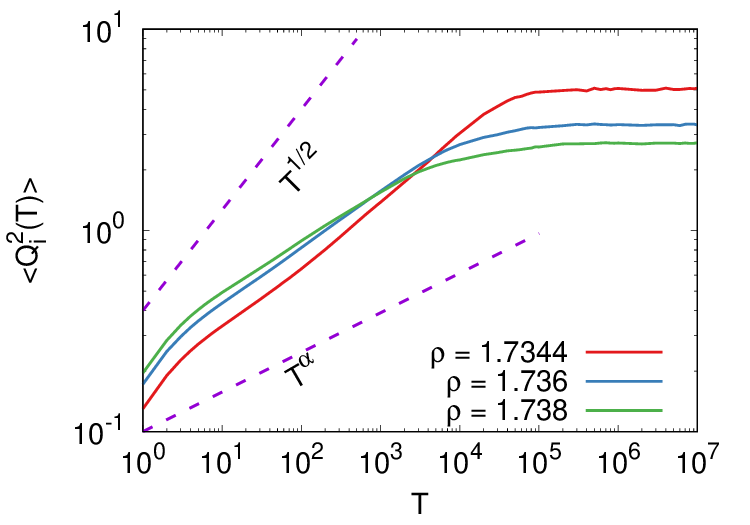}
    \caption{Integrated bond current fluctuation is plotted as a function of time near criticality for system size $L=5000$. The simulation data are plotted in solid lines for densities $\rhobar=1.7344$ (red line), $1.736$ (blue line), and $1.738$ (green line). The $T^\alpha$ guiding line is drawn using the theoretical estimation of the growth exponent $\alpha \simeq 0.197$, obtained using Eq.\eqref{eq:alpha_from_scaling_rel}. Another guiding line $T^{1/2}$ for subdiffusive is plotted to comparison.
}
    \label{fig:oslo_intj_fluc}
\end{figure}

In Fig. \ref{fig:oslo_intj_fluc}, we plot $\abr{\intc^2_i(T)}$ as a function of time for $\rhobar=1.7344$ (solid red line), $\rhobar=1.736$ (solid blue line) and $\rhobar=1.738$ (solid green line). All simulation data are taken for $L=5000$. A nontrivial crossover behavior that can be seen in this plot is that, although $\abr{\intc^2_i(T)}$ for $\rhobar=1.7344$ has the lowest intensity at the beginning ($T \ll 1$), eventually, the corresponding $\Sigma^2_Q(\Delta)$ becomes maximum for $T \gg 1$. This is followed by $\Sigma^2_Q(\Delta)$ for $\rhobar=1.736$ and then $\Sigma^2_Q(\Delta)$ for $\rhobar=1.738$. Indeed, we obtain by using Eq. \eqref{eq:oslo_new_fluctuation_relation} that $\Sigma^2_Q \sim \Delta^{-3/8}$ near criticality.
As a consequence, the initial time-dependent growth of cumulative current fluctuations is described by a power law with exponent $\alpha$,
\begin{align}
    \label{eq:oslo_current_fluctuation_asymptotic}
    \abr{\intc^2_i(T)} \sim T^\alpha,
\end{align}
where the exponent $\alpha$ can be estimated from Eq.\eqref{eq:oslo_new_fluctuation_relation}. As we approach criticality, $\abr{\intc^2_i(T)}$ must saturate at $T \simeq L^z$, where $z$ is the dynamic exponent. Using Eq. \eqref{eq:oslo_current_fluctuation_asymptotic}, the corresponding saturation value can be expressed as
\begin{align}
	\label{eq:Q2(T)_saturation_value}
	\lim_{T \rightarrow \infty} \abr{\intc^2_i(T, L)} \sim L^{\alpha z} 
	\sim \Delta^{-\delta},
\end{align}
where we use Eq.\eqref{eq:oslo_subsystem_mass_fluc}. Then, by using the finite-size-scaling  $L \simeq \Delta^{-\nu_\perp}$, we write the growth exponent $\alpha$ as
\begin{align}
	\label{eq:alpha_from_scaling_rel}
	\alpha = \frac{\delta}{z \nu_\perp}.
\end{align}
Now, by substituting the values of $\delta \simeq 3/8$, $z \simeq 10/7$ and $\nu_\perp \simeq 4/3$ \cite{Grassberger2016Oct}, we obtain $\alpha \simeq 63/320 \simeq 0.197$, shown in the guiding line $T^\alpha$ in Fig. \ref{fig:oslo_intj_fluc} (the other guiding line representing $T^{1/2}$ for subdiffusive growth, as observed in diffusive systems with a single conservation law, is provided for comparison purpose). Therefore, the near- and far-from-critical behaviors of time-integrated bond current fluctuation can be written in a combined form,
\begin{align}
\label{eq:oslo_exponent_alpha_def}
\abr{\intc_i^2(T)} \sim
\begin{cases}
T^{\alpha} & \alpha = {\delta}/{z \nu_\perp} \hspace{6pt} \text{for} \hspace{6pt} \rhobar \simeq \rho_c \\
\Sigma^2_{Q}(\rho) - {\rm const.} T^{\alpha} & \alpha = -{1}/{2} \hspace{6pt} \text{for} \hspace{6pt} \rhobar \gg \rho_c.
\end{cases}
\end{align}
Indeed, as we see later, the exponent $\alpha$ allows us to estimate corresponding exponents for the asymptotic behavior of the dynamic correlation function (and associated power spectra) for the instantaneous bond current as well as subsystem mass fluctuations (see Eqs.\eqref{eq:oslo_bond_corr_dyn}, \eqref{eq:oslo_ps_exponent} and \eqref{eq:oslo_sub_mass_ps_exponent} respectively).

\subsubsection{Instantaneous current correlation: Away from criticality}
\label{sec:oslo instantaneous_current_fluc}

\begin{figure}[!ht]
    \centering
    \includegraphics[width=1.0\linewidth]{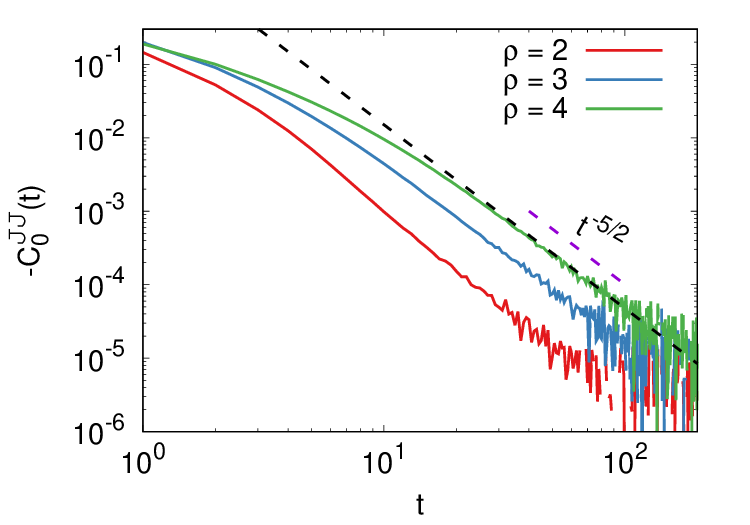}
    \caption{Far from  (above) criticality. The negative correlation of instantaneous bond current $-C_0^{\mathcal{JJ}}(t)$ is plotted vs. time for different away from criticality densities. The simulation data are plotted in solid lines for $\rhobar=2$ (red-colored line), $\rhobar=3$ (blue-colored line), $\rhobar=4$ (green-colored line). The corresponding theoretical asymptotic black-dotted line is plotted using the Eq.\eqref{eq:oslo_J0Jt_correlation-approx}  for $\rhobar=4.0$ and for system size $L=1000$. The asymptotic line excellently captures the $t^{-5/2}$ power-law decay of the correlation function; it should be contrasted with the conserved Manna sandpiles, for which the decay is $t^{-3/2}$.}
    \label{fig:oslo_JJ(tt)_corr}
\end{figure}

In this section, we demonstrate that the instantaneous current correlation decays faster than other diffusive systems, such as the Manna model. 
The instantaneous bond-current correlation, $C_r^{\mathcal{J} \mathcal{J}}(t)$ $=$ $\abr{\mathcal{J}_i(t)\mathcal{J}_{i+r}(0)}$ $-$ $\abr{\mathcal{J}_i(t)}\abr{\mathcal{J}_{i+r}(0)}$ for $t \geq \tp$, can be obtained by differentiating Eq.\eqref{eq:oslo_CrQQ(t,tp)-sol-full-expression} as
\begin{align}
    \label{eq:oslo_J0Jt_def}
    C_r^{\mathcal{J} \mathcal{J}}(t) = \sbr{\frac{d}{dt} \frac{d}{d\tp} 
    C_r^{\intc \intc}(t,\tp)}_{\tp = 0, t\geq 0},
\end{align}
which gives us,
\begin{align}
    \label{eq:oslo_J0Jt_correlation}
    C_r^{\mathcal{J} \mathcal{J}}(t) = \Gamma_r \delta(t)
    -a a^\prime \frac{1}{L} \sum_{q} e^{-\lambda_q a^\prime t} \frac{\lambda_q^2}{2} e^{-\imgi q r}.
\end{align}
The decay of the current correlation over time is governed by the second part of the right-hand side of Eq. \eqref{eq:oslo_J0Jt_correlation}. Its asymptotics for $t \gg 1$ and for a single bond $r=0$ can be understood by converting the sum into an integral in the continuum limit using $i \rightarrow x = i/L$, where $\lambda_q$ can be approximated as $\lambda_q \rightarrow \lambda(x) \simeq 4\pi^2 x^2$. Then, using the variable transformation $z=4\pi^2 x^2 a^\prime t$, we can write Eq. \eqref{eq:oslo_J0Jt_correlation} for $t \gg 1$ as follows:
\begin{align}
    \label{eq:oslo_J0Jt_correlation-approx}
    C_0^{\mathcal{J} \mathcal{J}}(t) \simeq 
    -\frac{a t^{-5/2}}{4 \pi  {a^\prime}^{3/2}} \int_0^\infty e^{-z} z^{3/2} dz = -\frac{3a t^{-5/2}}{16 \sqrt{\pi}  {a^\prime}^{3/2}}.
\end{align}
In Fig.\ref{fig:oslo_JJ(tt)_corr}, the negative instantaneous bond-current correlation
$-C_0^{\mathcal{JJ}}(t)$ is plotted as a function of time for different away from criticality densities. Simulation data are plotted as solid lines for $\rhobar=2$ (red colored line), $\rhobar=3$ (blue colored line) and $\rhobar=4$ (green colored line) for $L=1000$. The corresponding theoretical dotted line is plotted for $\rhobar=4.0$ using Eq.\eqref{eq:oslo_J0Jt_correlation-approx}, which indicates the $t^{-5/2}$ decay of correlation function in time.
The corresponding decay of the correlation function for other diffusive models with a single conserved quantity is contrastingly different from what we derived here and is given by $-t^{-3/2}$, which is much slower compared to the Oslo ricepile. We also determine the asymptotic behavior of $C^{\mathcal{J} \mathcal{J}}_r(t)$ using the dimensional analysis of $\abr{\intc_i^2(T)}$ and express it in terms of $\alpha$ as
\begin{align}
\label{eq:oslo_bond_corr_dyn}
C_0^{\mathcal{J} \mathcal{J}}(t) \sim -t^{-(2-\alpha)}.
\end{align}
Furthermore, the asymptotic behaviors near and away from criticality can be written using the $\alpha$ values from Eq. \eqref{eq:oslo_exponent_alpha_def} as
\begin{align}
    \label{eq:oslo_current_correlation_exponents}
    C_0^{\mathcal{J} \mathcal{J}}(t) \sim
    \begin{cases}
    -t^{-\br{2-{\delta}/{\nu_\perp z}}} & {\rm for} \hspace{6pt} \rhobar \simeq \rho_c, \\
        -t^{-5/2} & \hspace{1pt} {\rm for} \hspace{6pt} \rhobar \gg \rho_c,
    \end{cases}
\end{align}
where the asymptotic behavior in the far from critical (from above) regime is the same as that obtained in Eq.\eqref{eq:oslo_J0Jt_correlation-approx}.
Note  that, for the conserved Manna model \cite{Mukherjee2023Feb}, the decay of the near-critical current correlation function, $C_0^{\mathcal{J} \mathcal{J}}(t) \sim -t^{-(3/2+\mu)}$ with $\mu > 0$, is indeed faster compared to the far-from-critical one, where $\mu = 0$. This result implies a faster suppression of the dynamical fluctuation $\langle \intc_i^2 \rangle$ of bond current; in Ref. \cite{Mukherjee2023Feb}, we identified the phenomenon of a faster near-critical decay of the instantaneous current correlation function with $\mu > 0$ as ``dynamical hyperuniformity''.
However, for the Oslo model, the situation is qualitatively quite different. From Eq. \eqref{eq:oslo_current_correlation_exponents}, we observe that the corresponding values of the {\it dynamical hyperuniformity} exponents for near and away from criticality (from above) are $\mu = \br{1/2 - \delta / \nu_\perp z} \simeq 0.303$ and $\mu = 1$,  respectively.

The above results indicate that the decay of the instantaneous bond-current correlation away from criticality is faster than that near criticality. In other words, compared to the diffusive systems with a single conserved quantity, the Oslo model is dynamically more hyperuniform far from criticality than near criticality. The result underscores the significance of conserved quantities in determining large-scale fluctuations in these systems. As a consequence of the maximal (static as well as dynamic) hyperuniformity, which arises away from criticality, we also observe that the time-integrated correlation function for bond current away from criticality decays, in the thermodynamic limit $L \rightarrow \infty$, as
   \begin{align}
    \label{eq:oslo_Cj(t)-integrations}
    &\int_{-T}^T C_0^{\mathcal{J} \mathcal{J}}(t) dt = \frac{a(\rho)}{L}   \sum _q \lambda _q e^{-a'(\rho) T \lambda _q} \nonumber \\
&\simeq \frac{a }{2 \pi {a^\prime}^{3/2}}T^{-3/2} \int_0^\infty e^{-z} \sqrt{z} dz 
=    \frac{a}{4 \sqrt{\pi } {a^\prime}^{3/2}}T^{-3/2}.
   \end{align}
Perhaps not surprisingly, this particular decay behavior is considerably faster compared to the behavior of the same dynamic quantity for the conserved Manna sandpile \cite{Mukherjee2023Feb}, where, far from criticality, the time-integrated correlation was found to decay as $T^{-1/2}$.

\subsubsection{Power spectrum: Bond current}
\label{sec:oslo_ps}

Another way to characterize the dynamic fluctuation is to calculate the power spectrum. Theoretically, the power spectrum of the bond current in a steady state, denoted as $S_\mathcal{J}(f)$, can be obtained for a particular frequency $f$ by Fourier transforming the two-point correlation function $C^{\mathcal{J} \mathcal{J}}_r(t)$ given in Eq. \eqref{eq:oslo_J0Jt_correlation} with $r=0$ as follows,
\begin{align}
    \label{eq:oslo_SJ(f)_def}
    S_\mathcal{J}(f) = \int_{-\infty}^{\infty} dt C_0^{\mathcal{J} \mathcal{J}}(t) 
    e^{2 \pi \imgi ft}
    = \frac{a(\rhobar)}{L}
    \sum_q \frac{4 f^2 \pi^2 \lambda_q}{{a^\prime}^2 \lambda_q^2 + 4 f^2 \pi^2}.
\end{align} 
In the limit of the large system size, $L \gg 1$, the asymptotic behaviour of the above sum in the frequency domain $1/L^2 \ll f \ll 1$, can be obtained by approximating the above sum by the following integral,
\begin{align}
    \label{eq:oslo_SJ(f)_integral}
    S_\mathcal{J}(f) \simeq 2a \int_{\frac{1}{L}}^{\frac{1}{2}} dx \lambda(x) \frac{4\pi^2 f^2}{\lambda(x)^2 {a^\prime}^2 + 4\pi^2 f^2},
\end{align}
where in the continuum limit, we perform the substitution $q = 2\pi x$, which leads to $\lambda(x) \approx 4\pi^2 x^2$. Then by using the variable transformation, $x=\sqrt{f} y^{1/4} / \sqrt{2 \pi {a^\prime}}$, we convert the above integral into the following in the limit $L \rightarrow \infty$ and obtain
\begin{align}
\label{eq:oslo sjf asymptotic}
    S_\mathcal{J}(f) = f^{3/2} \frac{a \sqrt{\pi}}{\sqrt{2} {a^\prime}^{3/2}} 
    \int_0^\infty \frac{dy}{y^{1/4} (1+y)} = a \frac{\pi^{3/2}}{{a^\prime}^{3/2}} f^{3/2}.
\end{align}
By employing a straightforward dimensional analysis, one can obtain the exponent $\psi_{\mathcal{J}}$, governing the asymptotics $S_\mathcal{J}(f) \sim f^{\psi_{\mathcal{J}}}$, in terms of the growth exponent $\alpha$ of the time-integrated bond current, as shown below,
\begin{align}
    \label{eq:oslo_ps_exponent}
    \psi_{\mathcal{J}} = 1-\alpha.
\end{align}
Using the values of $\alpha$, given in Eq.\eqref{eq:oslo_exponent_alpha_def}, for near and away from criticality, we can write $\psi_{\mathcal{J}}$ as
\begin{align}
    \label{eq:oslo psi_j values}
    \psi_{\mathcal{J}} \sim
    \begin{cases}
     (1 - {\delta}/{\nu_\perp z}) \simeq 0.803 & {\rm for} \hspace{6pt} \rhobar \simeq \rho_c, \\
    {3}/{2} & {\rm for} \hspace{6pt} \rhobar \gg \rho_c, \\
\end{cases}
\end{align}
where away from criticality it is governed by the same exponent that appeared in Eq.\eqref{eq:oslo sjf asymptotic}.
Compared to diffusive systems with a single conserved quantity, where we have found $\psi_{\mathcal{J}} = 1/2+\mu$ \cite{Mukherjee2023Feb}, and where $\mu > 0$ suggests dynamical hyperuniformity, the Oslo model exhibits different behavior. Near criticality, $\mu = 1/2 - \delta/\nu_\perp z \simeq 0.303$ indicates dynamic hyperuniformity. On the other hand, away from criticality, $\mu = 1$, indicates the presence of maximal dynamic hyperuniformity in current fluctuation. This is because the steeper decay of the power spectrum with decreasing frequencies characterizes a faster vanishing rate of the current fluctuation.

In simulations, we compute the power spectrum by discretizing the instantaneous current into small temporal intervals, typically denoted as $\delta t$ \cite{Mukherjee2023Feb}. The Fourier transform is then calculated as follows,
\begin{align}
    \label{eq:oslo_J(t)dt_fourier}
    \tilde{\mathcal{J}}_{n;T} = \delta (t) \sum_{k=0}^{T-1} \mathcal{J}_i(k) e^{2\pi \iu f_n k},
\end{align}
where we denote $f_n=n/T$ for $T \gg 1$. Thus the power spectrum for bond current can be defined as
\begin{align}
\label{eq:oslo_SJ(f)_simulation}
S_{\mathcal{J}}(f_n) = \lim_{T \rightarrow \infty} \frac{1}{T} \langle \lvert\tilde{\mathcal{J}}_{n;T}\rvert^2 \rangle,
\end{align}
where $f_n = n/T$ for $T \gg 1$. The limit as $T$ tends to infinity represents an average over an infinitely long time period. In the limit os system size $L \rightarrow \infty$ large, the convergence of the discrete sum as given in Eq.\eqref{eq:oslo_SJ(f)_simulation} to its continuum limit is expected, as expressed in Eq.\eqref{eq:oslo_SJ(f)_def}.

\begin{figure}[!ht]
    \centering
    \includegraphics[width=1.0\linewidth]{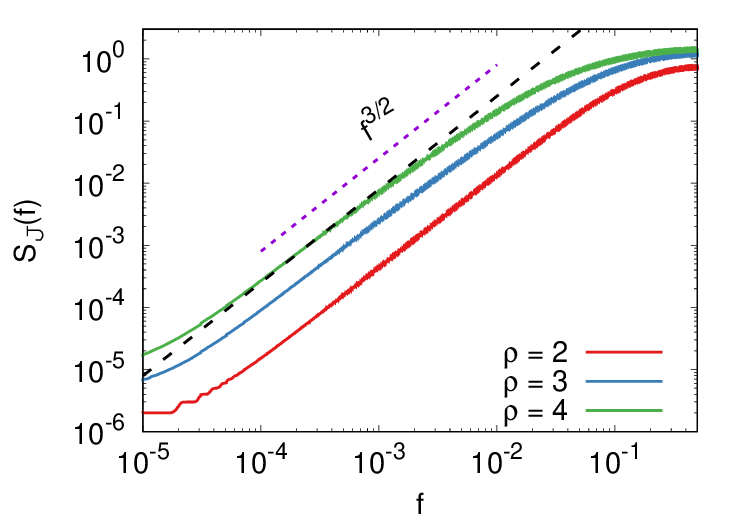}
    \includegraphics[width=1.0\linewidth]{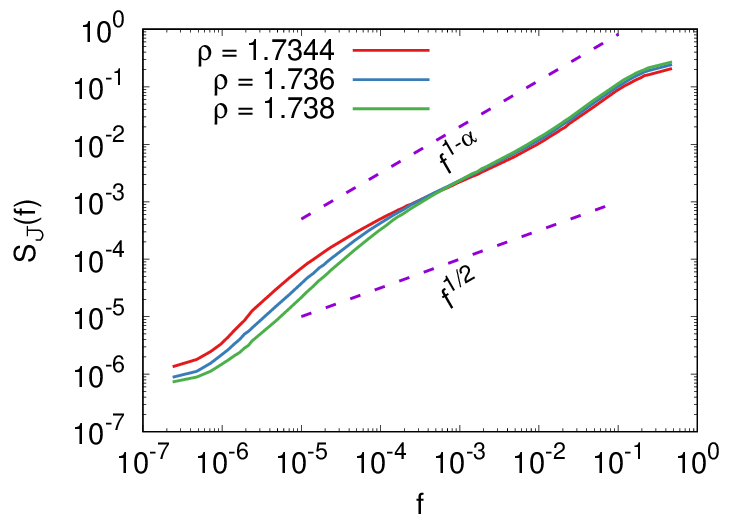}
    \caption{Power spectra near criticality. We plot power spectrum of the bond current as a function of frequency. \textit{Top panel:} This corresponds to the density regime away from criticality. Simulation data, taken with $L=1000$, are plotted as solid lines from bottom to top for $\rhobar=2$ (red), $\rhobar=3$ (blue), and $\rhobar=4$ (green). Black-dotted asymptotic line, plotted using Eq.\eqref{eq:oslo sjf asymptotic} for $\rhobar=4.0$ and nicely demonstrate the  $f^{\psi_{\mathcal{J}}}$, with $\psi_{\mathcal{J}} = 3/2$ decay of the power spectrum as $f \rightarrow 0$.
    \textit{Bottom panel:} This plot corresponds to the power spectrum of current to the density regime near criticality. Simulation data, taken with $L=5000$, are plotted as solid lines for densities $\rhobar=1.738$ (red), $1.736$ (blue), and $1.7344$ (green). The $f^{1-\alpha}$ guiding line demonstrates
our theoretical estimation of the decay of power spectrum near criticality and another $f^{1/2}$ guiding line is plotted for the comparison with random diffusive systems. The exponent $\psi_{\mathcal{J}} = 1 -\alpha \simeq 0.803$ is obtained by using Eq.\eqref{eq:oslo psi_j values}.
}
    \label{fig:oslo_ps}
\end{figure}

{In top panel} of Fig. \ref{fig:oslo_ps}, we plot far-from-critical power spectrum $S_{\mathcal{J}}(f)$ vs. frequency $f$ for densities $\rhobar =$ $2$ (red), $3$ (blue), and $4$ (green) in solid lines for $L=1000$. The corresponding theoretical result, obtained by using Eq. \eqref{eq:oslo sjf asymptotic}  for $\rhobar=4.0$, is shown in the plot as a black dashed line. This asymptotic theoretical result quite nicely captures the $f^{3/2}$ decay of the power spectrum as $f \to 0$.
\textit{In the bottom panel}, we present the simulation data of the power spectrum for near-critical densities $\rhobar = 1.7344$ (green line), $1.736$ (blue line), and $1.738$ (red line) for $L=5000$. The upper guiding line, $f^{1-\alpha}$, is drawn following the relation $\psi_{\mathcal{J}} = 1-\alpha$ as given in Eq. \eqref{eq:oslo_ps_exponent}, where the near-critical value $\psi_{\mathcal{J}} \simeq 0.803$ is provided in Eq.\eqref{eq:oslo psi_j values}. In the same figure, we also plot a lower guiding line, $f^{1/2}$, representing the normal decay of the power spectrum for a completely random diffusive system.

\subsubsection{Power spectrum: Subsystem mass}
\label{sec:oslo__mass_ps}

In this section, we compute the power spectrum of subsystem mass, given in Eq.\eqref{eq:oslo_subsystem_mass_Cr(t)_sol} by taking the Fourier transform of $C^{M_l M_l}(t)$, which can be expressed as given below,
\begin{align}
  \label{eq:oslo_subsys mass ps}
  S_{M}(f) = &\lim\limits_{T \rightarrow \infty}
  \int\limits_{-T}^{T} C^{M_l M_l}(t) e^{2\pi \imgi ft} dt \nonumber \\
  = &\frac{1}{L} \frac{a(\rho)}{2\actp(\rho)} \sum\limits_{q}
  \frac{2\lambda_q \actp(\rho)}{\lambda_q^2 {\actp}^2(\rho)+4\pi^2f^2}
  \lambda_{lq}.
\end{align}
Similarly, like the current power spectrum, this expression can also be written as an integral where we take the continuum limit $i \rightarrow x = i/L$ and the large system size limit $L \gg 1$. In the limit of the size of the large subsystem $l \gg 1$ and $l/L \ll 1$, we can also approximate $\lambda_{lq} \simeq 2$.
Furthermore, using the variable transformation, $x=\sqrt{f} y^{1/4} / \sqrt{2 \pi a^\prime}$, the mass power spectrum can finally be written as
\begin{align}
  \label{eq:oslo_subsys mass ps_continuum}
  S_{M}(f) &\simeq \frac{2a}{\actp} \frac{1}{4\sqrt{2} \pi ^{3/2} \sqrt{{\actp}} \sqrt{f}} \int_0^{\infty} \frac{dy}{y^{1/4} (1+y)} \nonumber \\
  &= \frac{a}{2 \sqrt{\pi } {\actp}^{3/2}} f^{-1/2}.
\end{align}
The asymptotic behavior, $S_{M}(f) \sim f^{-\psi_M}$, can also be obtained from the current power spectrum exponent $\psi_{\mathcal{J}}$, as we noted earlier in \cite{Mukherjee2023Feb}, the exponents $\psi_{\mathcal{J}}$ and $\psi_M$ follow the relation $\psi_M = 2 - \psi_{\mathcal{J}}$. This relationship is a consequence of mass conservation through diffusive time-evolution equation and holds true both away from and near criticality. Thus we obtain, by using Eq.\eqref{eq:oslo_ps_exponent}, the following scaling relation,
\begin{align}
\label{eq:oslo_sub_mass_ps_exponent}
\psi_M = 1+\alpha.
\end{align}
Immediately, using the Eq.\eqref{eq:oslo psi_j values}, we obtain the near and away from criticality values of $\psi_M$ as follows,
\begin{align}
\label{eq:oslo_psi_M}
    \psi_M =
    \begin{cases}
    (1+{\delta}/{\nu_\perp z}) \simeq 1.197 & {\rm for} \hspace{6pt} \rhobar \simeq \rho_c, \\
         {1}/{2} & {\rm for} \hspace{6pt} \rhobar \gg \rho_c ,
    \end{cases}
\end{align}
where the value of the away from criticality exponent is the same as that we obtained in Eq.\eqref{eq:oslo_subsys mass ps_continuum}.
That is, the power spectrum for subsystem mass in one dimension can be written as $S_{M}(f) \sim f^{-3/2 + \mu}$. Evidently, positive $\mu > 0$ corresponds to ``anomalous" (more suppressed than that in usual diffusive systems) fluctuation for subsystem mass; indeed the anomalous fluctuation arises from ``dynamic hyperuniformity" in the bond-current fluctuation. In the Oslo model, we obtain from Eq.\eqref{eq:oslo_psi_M}, $\mu = 1/2 - \delta / \nu_\perp z$ near criticality and $\mu=1$ away from criticality; these theoretical predictions are quite consistent with our estimates obtained in previous section in the context bond current fluctuation [see the text below eq. \eqref{eq:oslo_current_correlation_exponents}].

\begin{figure}[!ht]
    \centering
    \includegraphics[width=1.0\linewidth]{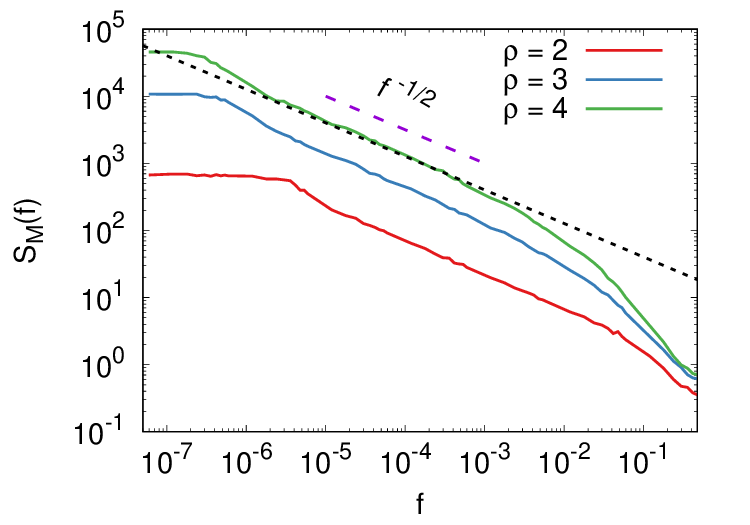}
    \includegraphics[width=1.0\linewidth]{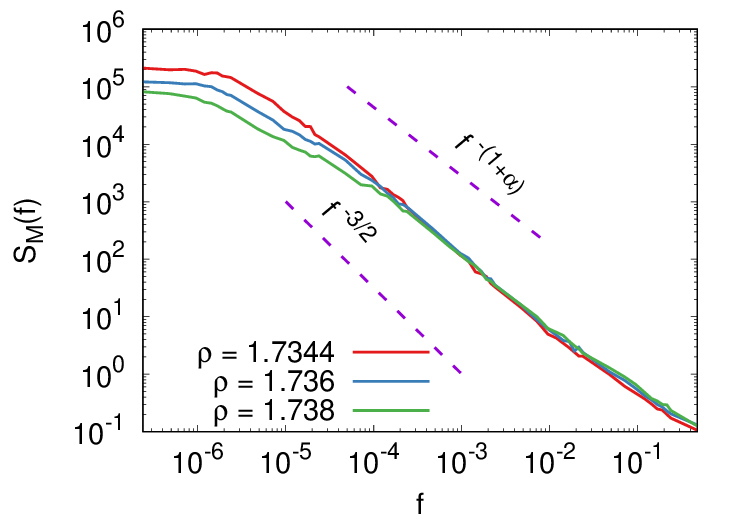}
    \caption{The power spectrum for subsystem mass is plotted against frequency. \textit{Top panel:} Simulation data for the power spectrum for subsystem size $l=500$ and system size $L=1000$ are plotted as a function of frequency for density values away from criticality, namely $\rhobar = 2$ (red), $3$ (blue), and $4$ (green) as solid lines, from bottom to top, respectively. The corresponding (asymptotic) black-dotted line is plotted using Eq.\eqref{eq:oslo_subsys mass ps_continuum} for $\rhobar=4.0$ and it confirms that, far from criticality, the power spectrum for subsystem mass grows as $f^{-1/2}$ in the small-frequency limit. \textit{bottom panel:} The power spectrum for subsystem mass, obtained from simulation for system size $L=5000$ and subsystem size $l=2500$, are plotted for near-critical density values $\rhobar = 1.7344$ (green line), $1.736$ (blue line) and $1.738$ (red line). The guiding line representing $f^{-(1+\alpha)}$ is the asymptotic power spectrum [obtained by using scaling theory in eqs. (\ref{eq:oslo_sub_mass_ps_exponent}) and (\ref{eq:oslo_psi_M})] for subsystem mass near criticality as $f \to 0$, whereas the subdiffusive guiding line representing $f^{-3/2}$  is provided for comparison purpose. 
    }
    \label{fig:oslo_mass_ps}
\end{figure}

In {top panel} of Fig. \ref{fig:oslo_mass_ps}, we plot subsystem-mass power spectrum, which is obtained for subsystem size $l=500$ and system size $L=1000$ as a function of the frequency $f$, represented by solid lines, for density values away from the criticality: $\rhobar = 2$ (red), $3$ (blue), and $4$ (green) from bottom to top. The corresponding asymptotic line for $\rhobar=4.0$ is plotted in black-dotted line using Eq.\eqref{eq:oslo_subsys mass ps_continuum}. This asymptotic line demonstrates a nice agreement between the simulation and the $f^{-1/2}$ growth of the power spectrum as $f \to 0$ away from criticality.
In {bottom panel}, we present similar data, but for different subsystem and system sizes $l=2500$ and $L=5000$, respectively, for near-critical densities $\rhobar = 1.7344$ (green line), $1.736$ (blue line), and $1.738$ (red line). The $f^{-(1+\alpha)}$ guiding line is drawn using Eq. \eqref{eq:oslo_sub_mass_ps_exponent}, with $\psi_M \simeq 1.197$. The lower guiding line of $f^{-3/2}$, which is for diffusive systems having a single conserved quantity, is plotted for comparison purposes only.

\section{Structure factor}
\label{sec:oslo_structure_factor}

The static density fluctuation can also be quantified by the structure factor $S(q) = \abr{\lvert \tilde{\delta m}_q(t) \rvert^2} / N$ \cite{Torquato2018Jun}, where $\tilde{\delta m}_q(t)$ is the Fourier transform of the excess mass $\delta m_i(t) = m_i(t) - \rhobar$ at a site $i$. For hyperuniform states, $S(q) \to 0$ as $q \to 0$, particularly, for the maximal hyperuniform states $S(q) \sim q^2$ in the small $q$ limit \cite{Hexner2017Jan}. In the following, we calculate $S(q)$ using the microscopic evolution equation for $\delta m_i(t)$,
\begin{align}
	\label{eq:mass_evl_eqn}
	\frac{\partial }{\partial t} \delta m_i(t) =& \br{\jd_{i-1} - \jd_i} + \br{\jfl_{i-1} - \jfl_i}, \nonumber \\
	\simeq& D(\rhobar) \Delta_{i,k} \delta m_k(t) + \br{\jfl_{i-1} - \jfl_i}.
\end{align}
We express $\delta m_i(t)$ and $\jfl_i(t)$ in the Fourier modes in the above equation and obtain,
\begin{align}
	\label{eq:mass_evl_eqn_fourier}
	\frac{\partial }{\partial t} \tilde{\delta m}_q(t) \simeq -D\lambda_q \tilde{\delta m}_q(t) + \br{e^{\imgi q} - 1} \tilde{\mathcal{J}}^{(fl)}_q(t),
\end{align}
where $\tilde{\mathcal{J}}^{(fl)}_q(t)$ is the Fourier transform of fluctuating current $\jfl$. Solving the above equation, we write $\tilde{m}_q(t)$ as
\begin{align}
	\label{eq:oslo_mass_fourier}
	\tilde{\delta m}_q(t) = \int_0^t d\tp e^{-\lambda_q D (t-\tp)} \br{e^{\imgi q} - 1} \tilde{\mathcal{J}}^{(fl)}_q(\tp).
\end{align}
Now, we calculate the structure factor $S(q)$, using the $\tilde{\delta m}_q(t)$ as
\begin{align}
	\label{eq:structure_factor}
	&S(q) =  \frac{1}{L\rhobar} \abr{\lvert \tilde{\delta m}_q(t) \rvert^2}\nonumber \\ &= \frac{1}{L\rhobar} \int_0^t d\tp \int_0^t d\tpp e^{-\lambda_q D (\tp-\tpp)} \lambda_q \abr{\tilde{\mathcal{J}}^{(fl)}_q(\tp) \tilde{\mathcal{J}}^{(fl)}_{-q}(\tpp)}.
\end{align}
Using the correlation of function of $\jfl$, given in Eq.\eqref{eq:oslo fl current corr sol}, we can write,
\begin{align}
\abr{\tilde{\mathcal{J}}^{(fl)}_q(\tp) \tilde{\mathcal{J}}^{(fl)}_{q^\prime}(\tpp)} = L a(\rhobar) \lambda_q \delta_{q,-q^\prime} \delta(\tp-\tpp),
\end{align}
which by putting in Eq.\eqref{eq:structure_factor}, we obtain,
\begin{align}
\label{eq:final_S_q(t)}
S(q) = \br{1 - e^{-2\lambda_q D t}} \frac{\lambda_q}{2D\rhobar} a(\rhobar).
\end{align}
Finally, in the limit $t \to \infty$ and approximating $\lambda(q) \simeq q^2$ for $q \to 0$, we obtain the steady-state structure factor
\begin{align}
	\label{eq:final_S(q)}
	S(q) \simeq \frac{a(\rhobar)}{2D\rhobar} q^2,
\end{align}
the expression, which is valid far from criticality.

\begin{figure}[!ht]
    \includegraphics[width=1.0\linewidth]{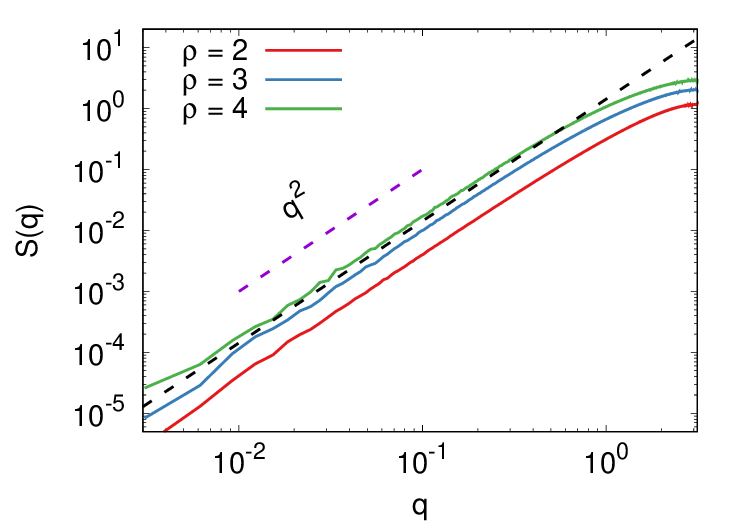}
    \includegraphics[width=1.0\linewidth]{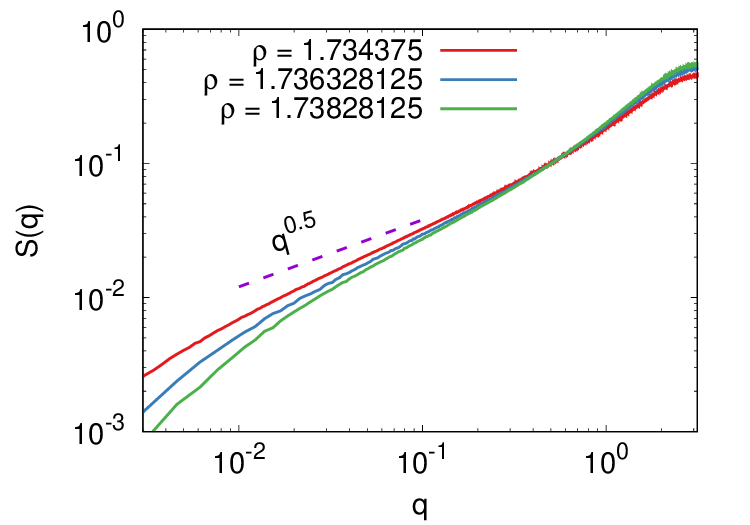}
    \caption{
    \textit{In the top panel} we plot the structure factor $S(q)$ for away from critical density values as a function of $q$ for a system size of $L=2^{11}$. We represent the simulation data with solid lines for density values of $\rhobar=2$, $3$, and $4$ from bottom to top. The dotted line corresponds to the theoretical asymptotic $S(q)$ for $\rhobar = 4$, plotted using Eq. \eqref{eq:final_S(q)}. \textit{In the bottom panel} we plot $S(q)$ for near critical density value of  $\rhobar = 1.734375$ for $L=2^{14}$, and of densities $\rhobar=1.736328125$, $1.73828125$ for $L=2^{13}$. In addition, we include the guideline $q^{0.5}$ to denote the structure factor's dependence on small $q$ values close to criticality. The hyperuniform fluctuation of subsystem mass can be used to determine this dependence (see Eq. \eqref{eq:oslo_hyperuniformity}).
}
    \label{fig:s(q)}
\end{figure}

In the top panel of Fig. \ref{fig:s(q)}, we plot the structure factors obtained from simulations for $L=2^{11}$, representing densities away from critical values: $\rhobar=2$ (red), $\rhobar=3$ (blue), and $\rhobar=4$ (green) shown as solid lines. Additionally, we plot our asymptotic expression in Eq. \eqref{eq:final_S(q)} as a black dotted line for $\rhobar=4$, demonstrating good agreement with the simulation data. The dotted magenta line represents the $q^{2}$ guiding line, denoting the functional dependence of the structure factor away from criticality on $q$. In the bottom panel, we plot the structure factor for densities near criticality: $\rhobar = 1.734375$ for $L=2^{14}$, and densities $\rhobar=1.736328125$ and $1.73828125$ for $L=2^{13}$. We also plot the guiding line $q^{0.5}$ in dotted magenta line to signify the dependence of the structure factor on small $q$ values near criticality. This dependence can be derived from the hyperuniform fluctuation of subsystem mass (see Eq. \eqref{eq:oslo_hyperuniformity}).

\section{Tagged particle diffusion}
\label{sec:oslo_tagged_particle}

\begin{figure}[!ht]
    \centering
    \includegraphics[width=1.0\linewidth]{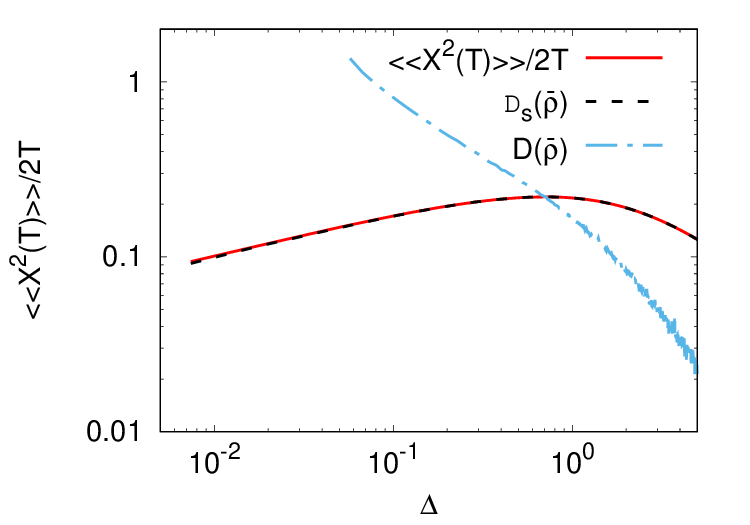}
    \caption{ Mean-square fluctuation of tagged particle displacement up to time $T$ (solid red colored line) vs. shifted density $\Delta$, where double angular brackets $\abr{\abr{X^2(T)}} = \sum_\alpha \abr{X_\alpha^2(T)}/N$ imply average over both particles and trajectories. Simulations (solid red line) and the theoretically obtained self-diffusion coefficient $\mathcal{D}_s(\rhobar)$ (dashed black colored line) as in Eq.\eqref{eq:olso_self-diffusivity-form} provides a very good agreement. We also plot the bulk-diffusion coefficient $D(\rhobar)=a'(\rhobar)$ vs. $\Delta=\rhobar-\rho_c$ (dot-dashed blue colored line), by calculating $D(\rhobar)$ through direct simulations where we use our theoretical result as in Eq.\eqref{D-rho}. Note that the bulk-diffusion coefficient and self-diffusion coefficient have qualitatively quite different behaviors. }
    \label{fig:oslo_self_diffusivity}
\end{figure}

In this section, we examine the variance of the displacements $X_\alpha(T)$ of a tagged particle $\alpha$ in a temporal domain $[0,T]$. Since a particle can only hop a distance of $+1$ or $-1$ from the toppled site, the variance of the total hop length $\langle X^2_\alpha(T) \rangle$ depends solely on the number of topplings experienced by the tagged particle within this time interval, denoted as $N_\alpha^{(h)}(T)$, as shown in the following equation,
\begin{align}
    \label{eq:oslo_X2(T)-N(T)}
    \langle X_{\alpha}^2(T) \rangle = N_\alpha^{(h)}(T).
\end{align}
By summing over all the tagged particles of the system, the total variance can be written in terms of twice of the total topplings $N^{(tp)}(T)$ during that interval as
\begin{align}
    \label{eq:oslo_total_X2(T)-N(T)}
    \sum_\alpha \langle X_{\alpha}^2(T) \rangle = \sum_\alpha N_\alpha^{(h)}(T) = 2 N^{(tp)}(T),
\end{align}
as at each toppling two particle jumps out of the site.
The number of total toppling in the system, $N^{(tp)}(T)$, on average, is equal to the number of active site density $a(\rhobar)$ times corresponding spacetime volume,
\begin{align}
    \label{eq:oslo_N(tp)}
    N^{(tp)}(T) = a(\rhobar) LT.
\end{align}
Due to the homogeneity of the system, we can express the left-hand side of Equation \eqref{eq:oslo_total_X2(T)-N(T)} as the total number of particles multiplied by the variance of a particle tagged $\abr{X^2(T)}$. Using Equation \eqref{eq:oslo_N(tp)}, we can rewrite Equation \eqref{eq:oslo_total_X2(T)-N(T)} as follows,
\begin{align}
    \label{eq:oslo_total_X2(T)-a(rho)}
    N \langle X^2(T) \rangle = 2 a(\rhobar) LT \hspace{6pt} \textit{or} \hspace{6pt} 
    \langle X^2(T) \rangle = 2 \frac{a(\rhobar)}{\rhobar} T.
\end{align}
The self-diffusion coefficient $\mathcal{D}_s(\rhobar)$ in the steady state is defined via the time-dependent (steady-state) mean-squared displacement $\abr{X^2(T)}$  as
\begin{align}
    \label{eq:oslo_self_diffusivity}
    \abr{X^2(T)} = 2 \mathcal{D}_s(\rhobar) T,
\end{align}
comparing this with Eq.\eqref{eq:oslo_total_X2(T)-a(rho)}, we obtain the exact expression of self-diffusivity in terms of the activity and the corresponding global density $\rhobar$ as
\begin{align}
    \label{eq:olso_self-diffusivity-form}
    \mathcal{D}_s(\rhobar) = \frac{a(\rhobar)}{\rhobar}.
\end{align}
The mean square fluctuation of the cumulative displacement of the tagged particles up to time $T$ (represented by the solid red line) is plotted in Fig. \ref{fig:oslo_self_diffusivity} as a function of the shifted density $\Delta$.
We denote double averaging over both trajectories and particles by $\abr{\abr{X^2(T)}} = \sum_\alpha \abr{X_\alpha^2(T)}/N$.
We observe excellent congruence between the simulation data of $\abr{\abr{X^2(T)}}/2T$ (solid red line) and the self-diffusion coefficient $\mathcal{D}_s(\rhobar)$ obtained theoretically in Eq. \eqref{eq:olso_self-diffusivity-form} (shown as the dashed black line).
Furthermore, to emphasize the contrasting behavior between the bulk- and self-diffusion coefficient, we display $D(\rhobar) = a'(\rhobar)$ in the same figure (depicted by dashed-dotted blue colored line), utilizing the relation in Eq.\eqref{D-rho}.
Particularly near criticality, we observe that the activity and consequently the self-diffusion coefficient approach zero as the global density approaches its critical value. Meanwhile, the self-diffusion coefficient diverges, indicating anomalous transport. However, away from criticality, both coefficients tend towards zero in different manners, as depicted in the figure.
We note that the self-diffusion coefficient for the conserved Manna sandpile satisfies a relationship similar to that given in Eq. \eqref{eq:olso_self-diffusivity-form}.

\section{Summary and conclusions}
\label{eq:oslo_summary}

In this paper, we use a microscopic approach for investigating the one-dimensional Oslo model on a ring geometry having $L$ discrete sites. We provide a large-scale characterization of various static and dynamic properties of mass and current in the system near as well as far from (above) criticality. The model is a paradigm for out-of-equilibrium systems, that conserve both mass and center-of-mass (CoM), but lack time-reversal symmetry (i.e., detailed balance is violated). We show that, despite having highly constrained microscopic dynamics due to CoM conservation, the Oslo model in fact exhibits {\it diffusive} relaxation far from criticality and {\it superdiffusive} relaxation near criticality.
In the near-critical scaling regime, the relaxation time has the following algebraic dependence on system size $L$, i.e., $\tau_r \sim L^z$, where the dynamical exponent $z = 2 - (1-\beta)/\nu_{\perp} < 2$ is determined by the two static exponents $\beta$ and $\nu_{\perp}$. Indeed the above scaling relation satisfied by the dynamic exponent is in perfect agreement with the near-rational values of the exponents, which were obtained through large-scale simulations in Ref. \cite{Grassberger2016Oct}.

Quite interestingly, in the long-time limit, the additional CoM conservation manifests itself in making the temporal growth rate of time-integrated current fluctuation, and consequently, {\it the particle mobility, to vanish}
In other words, current fluctuations become anomalously suppressed as compared to that observed in diffusive systems with only mass conservation, leading to hyperuniformity in spatial as well as in temporal domains in the system. 
To gain a better understanding of dynamic fluctuations, we implement a closure scheme, which helps one to calculate unequal-time (two-point) correlation functions, and the associated power spectra, involving current and mass. We obtain a mass-conservation principle as encoded in eq. \eqref{eq:oslo_new_fluctuation_relation}, which connects (dynamic) current and (static) mass fluctuations and exactly determines the decay exponents of the respective dynamic correlation functions near criticality in terms of the standard static exponents. We calculate the decay exponent exactly within the closure scheme, away from criticality. In the far-from-critical regime, we also analytically calculate the static structure factor, which has the following behaviour $S(q) \sim q^2$ in the limit of small wave number $q\to 0$ [see eq. \eqref{eq:final_S(q)}]. Quite strikingly, the static structure factor far from criticality exhibits class I hyperuniformity \cite{Torquato2021Nov}, whereas it exhibits class III hyperuniformity near criticality.

\begin{table}[!ht]

\centering
\begin{tabular}{|cc|cc|cc|}
\hline
\multicolumn{2}{|c|}{\textbf{\textit{Observables}}}                         & \multicolumn{2}{c|}{\textbf{\textit{Near criticality}}} & \multicolumn{2}{c|}{\textbf{\textit{Away from criticality}}} \\ \hline
\multicolumn{1}{|c|}{$\abr{\intc^2_i(T)}$} &
  $T^{\frac{1}{2} - \mu}$ &
  \multicolumn{1}{c|}{\textbf{\textit{Manna}}} &
  \textbf{\textit{Oslo}} &
  \multicolumn{1}{c|}{\textbf{\textit{Manna}}} &
  \textbf{\textit{Oslo}} \\ \hline
\multicolumn{1}{|c|}{$S_{\mathcal{J}}(f)$} &
  $f^{\frac{1}{2} + \mu}$ &
  \multicolumn{1}{c|}{\multirow{2}{*}{$\mu = \frac{\beta+1}{2\nu_\perp z}$}} &
  \multirow{2}{*}{$\mu=\frac{1-2\delta}{2 \nu_\perp z}$} &
  \multicolumn{1}{c|}{\multirow{2}{*}{$\mu=0$}} &
  \multirow{2}{*}{$\mu=1$} \\ \cline{1-2}
\multicolumn{1}{|c|}{$S_M(f)$} & $f^{-\frac{3}{2} + \mu}$ & \multicolumn{1}{c|}{}        &        & \multicolumn{1}{c|}{}          &           \\ \hline
\end{tabular}

\begin{tabular}{|l|ll|ll|}
\hline
\multirow{2}{*}{\begin{tabular}[c]{@{}l@{}}Transport \\ coefficients\end{tabular}} & \multicolumn{2}{l|}{Near criticality}                        & \multicolumn{2}{l|}{Away from criticality}               \\ \cline{2-5} 
                                                                                   & \multicolumn{1}{l|}{Manna}              & Olso               & \multicolumn{1}{l|}{Manna}            & Oslo             \\ \hline
$\chi$                                                                             & \multicolumn{1}{l|}{$\Delta^\beta$}     & $0$                & \multicolumn{1}{l|}{$2a$}             & $0$              \\ \hline
$D$                                                                                & \multicolumn{1}{l|}{$\Delta^{\beta-1}$} & $\Delta^{\beta-1}$ & \multicolumn{1}{l|}{$a^\prime(\rho)$} & $a^\prime(\rho)$ \\ \hline
$\mathcal{D}_s$                                                                    & \multicolumn{1}{l|}{$\Delta^\beta$}     & $\Delta^\beta$     & \multicolumn{1}{l|}{$a(\rho)/\rho$}   & $a(\rho)/\rho$   \\ \hline
\end{tabular}

\caption{\label{table:conclusion_data_set} A comparison of the dynamic fluctuation properties of time-integrated bond-current ${\cal Q}_i(T)$ in time interval $\in [0,T]$, and the power-spectra $S_{\cal J}(f)$ and $S_M(f)$ of the instantaneous bond current and subsystem-mass, respectively, and the various density-dependent transport coefficients - the bulk-diffusion coefficient $D(\rho)$, the mobility $\chi(\rho)$ and the self-diffusion coefficient ${\cal D}_s(\rho)$, in the Oslo and the Manna models. The mobility $\chi = \lim_{T/L^2 \to \infty, L \to \infty} L \abr{\intc^2_i(T,L,\Delta)}/T$ is defined by first taking the infinite-time limit ($T \to \infty$) and then the infinite-volume limit ($L \to \infty$).
}

\end{table}

Notably, the dynamic properties of the Oslo model are qualitatively different from that observed in diffusive systems with a single conservation law, e.g., symmetric simple exclusion processes \cite{Sadhu2016Nov}, and thus from that in the conserved Manna sandpiles \cite{Mukherjee2023Feb}. We provide a comparison between the Oslo model and the Manna sandpile (conserved-mass versions) in Table \ref{table:conclusion_data_set}, to emphasize concisely the similarities and differences in the large-scale temporal structures of the two models. It is worth mentioning that the results obtained here are not specific to the Oslo model, but should be valid for a broad class of models with both mass and CoM conservation.

In summary, when compared to other recently studied dynamically constrained systems, such as those with dipole-moment conservation \cite{Feldmeier2020Dec, Morningstar2020Jun, Han2023Apr}, we have presented a prototypical CoM conserving model with a contrasting relaxation mechanism, implying far richer phenomenological structure of these systems than that anticipated earlier. 
Most crucially, our findings underline the significance of time-reversal symmetry (or the lact of it) in determining the large-scale dynamical structure of such systems.
We believe they would elucidate the dynamical origin of anomalous relaxation and hyperuniform fluctuations in systems with multiple conservation laws, and provide a fresh perspective on the general theoretical understanding of the problem.

\section*{Acknowledgement}

We thank Deepak Dhar and S. S. Manna for very helpful discussions. We thank Deepak Dhar for careful reading of the manuscript.

\bibliography{references}

\newpage

\appendix
\widetext

\section{Equal-time mass and integrated current correlation function}
\label{sec:oslo_appendix_mQ(t,t)-update-rules}

To obtain the Eq. \eqref{eq:oslo_mQ(t,t)-evl-eqn},
the update rules of the function $m_i(t)\intc_{i+r}(t)$ given as
\begin{align}
    \label{eq:oslo_mQ(t,t)-update-rules}
    m_i(t+dt) \intc_{i+r}(t+dt) = 
    \begin{cases}
        \textit{\textbf{events}} & \textit{\textbf{probabilities}} \\
        \br{m_i(t) + 1} \br{\intc_{i+r}(t) + 1} & \hata_{i+1} \delta_{i+r,i+1} dt \\
        \br{m_i(t) + 1} \br{\intc_{i+r}(t) - 1} & \hata_{i+1} \delta_{i+r,i} dt \\
        \br{m_i(t) + 1} \br{\intc_{i+r}(t) + 1} & \hata_{i-1} \delta_{i+r,i-1} dt \\
        \br{m_i(t) + 1} \br{\intc_{i+r}(t) - 1} & \hata_{i-1} \delta_{i+r,i-2} dt \\
        \br{m_i(t) - 2} \br{\intc_{i+r}(t) + 1} & \hata_{i} \delta_{i+r,i} dt \\
        \br{m_i(t) - 2} \br{\intc_{i+r}(t) - 1} & \hata_{i} \delta_{i+r,i-1} dt \\
        \br{m_i(t) + 1} \intc_{i+r}(t) & \hata_{i+1} \br{1-\delta_{i+r,i+1}-\delta_{i+r,i}} dt \\
        \br{m_i(t) + 1} \intc_{i+r}(t) & \hata_{i-1} \br{1 - \delta_{i+r,i-1} - \delta_{i+r,i-2}} dt\\
        \br{m_i(t) - 2} \intc_{i+r}(t) & \hata_i \br{1 - \delta_{i+r,i} - \delta_{i+r,i-1}} dt \\
        m_i(t) \br{\intc_{i+r}(t) + 1} & \hata_{i+r}  \\
        & \br{1 - \delta_{i+r,i} - \delta_{i+r+1,i} - \delta_{i+r-1,i}} dt \\
        m_i(t) \br{\intc_{i+r}(t) - 1} & \hata_{i+r+1}  \\
        & \br{1 - \delta_{i+r,i} - \delta_{i+r+1,i} - \delta_{i+r+2,i}} dt \\
        m_i(t) \intc_{i+r}(t) & 1-\Sigma dt,
    \end{cases}
\end{align}
where $\Sigma dt$ is the sum of probabilities of all previous events. This update rules gives us the following evolution equation of the correlation of equal-time mass and current,
\begin{align}
    \label{eq:oslo_m(t)Q(t)-evl-eqn}
    \frac{d}{dt} \abr{m_i(t) \intc_{i+r}(t)} = \Delta_{i,k} \abr{\hata_k(t) \intc_{i+r}(t)} +f_{i,r}(t);
\end{align}
the source term $f_{i,r}(t)$ has the following representation,
\begin{align}
    \label{eq:oslo_m(t)Q(t)-evl-eqn-source}
    f_{i,r}(t) =& \abr{m_i(t) \hata_{i+r}(t)} - \abr{m_i(t) \hata_{i+r+1}(t)} + \nonumber \\
    &\abr{\hata_{i+1}} \br{\delta_{i+r,i+1} - \delta_{i+r,i}} + \abr{\hata_{i-1}} \br{\delta_{i+r,i-1} - \delta_{i+r,i-2}} -2
    \abr{\hata_i}\br{\delta_{i+r,i} - \delta_{i+r,i-1}},
\end{align}
and in the steady state, it will simply be,
\begin{align}
    \label{eq:oslo_m(t)Q(t)-evl-eqn-source-steady-state}
    f_{r}(t) =& C^{m\hata}_r(t,t) - C^{m\hata}_{r+1}(t,t) + a \cbr{3\br{\delta_{0,r+1} - \delta_{0,r}} + \br{\delta_{0,r-1} - \delta_{0,r+2}}}.
\end{align}
This completes the derivation of Eq.\eqref{eq:oslo_mQ(t,t)-evl-eqn}.

The corresponding correlation function $C^{m \hata}_r(t,t)$ is derived in the following section.

\section{Equal-time mass-mass correlation}
\label{sec:oslo_appendix_equal-time-mass-mass_correlation}

To obtain Eq. \eqref{eq:oslo_m(t)a(t)-sol}, we write
the evolution equation of the equal-time and unequal space mass-mass correlation function using the following update rules,
\begin{align}
    \label{eq:oslo_mm(tt)_update_eqn}
    &m_i(t+dt) m_{i+r}(t+dt) = \nonumber \\
    &\begin{cases}
        \textbf{\textit{events}} &\textbf{\textit{probabilities}} \\
        \br{m_i(t)-2} \br{m_{i+r}(t)-2} & \hata_i \delta_{i,i+r} dt \\
        \br{m_i(t)-2} \br{m_{i+r}(t)+1} & \hata_i \delta_{i+1,i+r}dt \\
        \br{m_i(t)-2} \br{m_{i+r}(t)+1} & \hata_i \delta_{i-1,i+r}dt \\
        \br{m_i(t)+1} \br{m_{i+r}(t)+1} & \hata_{i-1} \delta_{i,i+r}dt \\
        \br{m_i(t)+1} \br{m_{i+r}(t)-2} & \hata_{i-1} \delta_{i-1,i+r}dt \\
        \br{m_i(t)+1} \br{m_{i+r}(t)+1} & \hata_{i-1} \delta_{i-2,i+r}dt \\
        \br{m_i(t)+1} \br{m_{i+r}(t)+1} & \hata_{i+1} \delta_{i,i+r}dt \\
        \br{m_i(t)+1} \br{m_{i+r}(t)-2} & \hata_{i+1} \delta_{i+1,i+r} dt \\
        \br{m_i(t)+1} \br{m_{i+r}(t)+1} & \hata_{i+1} \delta_{i+2,i+r}dt \\
        \br{m_i(t)-2} m_{i+r}(t) & \hata_i \br{1-\delta_{i+1,i+r}-\delta_{i,i+r}-\delta_{i-1,i+r}}dt \\
        \br{m_i(t)+1} m_{i+r}(t) & \hata_{i-1} \br{1-\delta_{i-1,i+r}-\delta_{i,i+r}-\delta_{i-2,i+r}}dt \\
        \br{m_i(t)+1} m_{i+r}(t) & \hata_{i+1} \br{1-\delta_{i+1,i+r}-\delta_{i,i+r}-\delta_{i+2,i+r}}dt \\
        m_i(t) \br{m_{i+r}(t)-2} & \hata_{i+r} \br{1-\delta_{i,i+r+1}-\delta_{i,i+r} - 
        \delta_{i,i+r-1}}dt \\
        m_i(t) \br{m_{i+r}(t)+1} & \hata_{i+r+1} \br{1-\delta_{i,i+r+1}-\delta_{i,i+r+2} - 
        \delta_{i,i+r}}dt \\
        m_i(t) \br{m_{i+r}(t)+1} & \hata_{i+r-1} \br{1-\delta_{i,i+r-1}-\delta_{i,i+r} - 
        \delta_{i,i+r-2}} dt \\
        m_i(t) m_{i+r}(t) & 1-\Sigma dt,
    \end{cases}
\end{align}
where $\Sigma dt$ is the sum of probabilities of all previous events. The corresponding evolution equation of $C^{mm}_r(t,t)$ can be written using the above update rules as
\begin{align}
    \label{eq:oslo_mm(tt)_evl_eqn}
    \frac{d}{dt} C^{mm}_r(t,t) = \sum_k\Delta_{i,k} \abr{\hata_k m_{i+r}} +
    \sum_k\Delta_{i+r,k} \abr{m_i \hata_{k}} + B_{i,i+r},
\end{align}
where $B_{i,i+r}$ is the source part of this correlation, given as
\begin{align}
    \label{eq:olso_Bir_source}
    B_{i,i+r}=& \delta_{i,i+r} \br{4\hata_i + \hata_{i-1} + \hata_{i+1}} - 
    2\delta_{i-1,i+r} \br{\hata_i + \hata_{i-1}} - 2\delta_{i+1,i+r} \br{\hata_i + \hata_{i+1}} +
    \nonumber \\
    &\delta_{i-2,i+r} \hata_{i-1} + \delta_{i+2,i+r} \hata_{i+1}.
\end{align}
In the steady state we must have $\frac{d}{dt} C^{mm}_r(t,t) = 0$ and using the translation symmetry, Eq.\eqref{eq:oslo_mm(tt)_evl_eqn} can be written as
\begin{align}
    \label{eq:oslo_sdst_mm_evl_eqn}
    2 \br{C^{\hata m}_{r-1}-C^{\hata m}_{r}+C^{\hata m}_{r+1}} + B_r = 0,
\end{align}
and the source term $B_r$ is given by,
\begin{align}
    \label{eq:oslo_sdst_mm_source}
    B_r = 6 a(\rhobar) \delta_{0,r} - 4a(\rhobar) \br{\delta_{0,r+1}+\delta_{0,r-1}} + 
    a(\rhobar) \br{\delta_{0,r+2}+\delta_{0,r-2}}.
\end{align}
Note that we can also straightforwardly derive Eq. \eqref{eq:oslo_Cr(mm)_eqn} in the main text simply by inserting our truncation relation, as given in Eq. \eqref{eq:oslo_current_approximation}, into Eq. \eqref{eq:oslo_sdst_mm_evl_eqn}.
We can solve Eq.\eqref{eq:oslo_sdst_mm_evl_eqn} by multiplying both sides by $z^r$ and defining the generating function $G(z) = \sum_{r=0}^{\infty} z^r C^{\hata m}_r$. Imposing the convergence of $G(z) < \infty$ when $z < \infty$, we can write the generating function as
\begin{align}
    \label{eq:oslo_mass_mass_generating function}
    G(z) = a(\rhobar) - \frac{a(\rhobar)}{2} z.
\end{align}
From the generating function of above, we write the correlation function $C^{\hata m}_r$ as
\begin{align}
    C^{\hata m}_r = a(\rhobar) \delta_{0,r} - \frac{a}{2} \br{\delta_{r+1} + \delta_{r-1}}, 
\end{align}
and thus we prove Eq. \eqref{eq:oslo_m(t)a(t)-sol}.

\section{Equal-time current-current correlation}
\label{sec:oslo_appendix_equal-time-current-current_correlation}

To derive Eq. \eqref{eq:oslo_QQtt_evl_eqn}, we write the
evolution equation of the equal-time unequal-space correlation of integrated current using the following update equation,
\begin{align}
    \label{eq:oslo_QQtt_update_eqn}
    \intc_i(t+dt) \intc_{i+r}(t+dt) = 
    \begin{cases}
        \textbf{\textit{events}} & \textbf{\textit{probabilities}} \\
        \br{\intc_i(t) + 1} \br{\intc_{i+r}(t) + 1} & \hata_i(t) \delta_{i,{i+r}} dt \\
        \br{\intc_i(t) + 1} \br{\intc_{i+r}(t) - 1} & \hata_i(t) \delta_{i-1,{i+r}} dt \\
        \br{\intc_i(t) - 1} \br{\intc_{i+r}(t) - 1} & \hata_{i+1}(t) \delta_{i,{i+r}} dt \\
        \br{\intc_i(t) - 1} \br{\intc_{i+r}(t) + 1} & \hata_{i+1}(t) \delta_{i+1,{i+r}} dt \\
        \br{\intc_i(t) + 1} \intc_{i+r}(t) & \hata_{i}(t) \br{1 - \delta_{i,{i+r}} - 
        \delta_{i-1,{i+r}}}dt \\
        \br{\intc_i(t) - 1} \intc_{i+r}(t) & \hata_{i+1}(t) \br{1 - \delta_{i,{i+r}}-
        \delta_{i+1,{i+r}}}dt \\
        \intc_i(t) \br{\intc_{i+r}(t) + 1} & \hata_{{i+r}}(t) \br{1 - \delta_{i,{i+r}}-
        \delta_{i,{i+r-1}}}dt \\
        \intc_i(t) \br{\intc_{i+r}(t) - 1} & \hata_{{i+r}+1}(t) \br{1 - \delta_{i,{i+r}}-
        \delta_{i,{i+r+1}}}dt \\
        \intc_i(t) \intc_{i+r}(t) & 1 - \Sigma dt,
    \end{cases}
\end{align}
where $\Sigma dt$ is the probability of happening nothing in the time interval $dt$. The corresponding dynamical equation can be written as
\begin{align}
    \label{eq:oslo_QQ(tt)_evl_eqn}
    \pd{t} C^{\intc\intc}_r(t,t) = \Gamma_{i,i+r}(t) + \abr{\jd_i(t) \intc_{i+r}(t)}_c +
    \abr{\intc_i(t) \jd_{i+r})(t)}_c,
\end{align}
where $\Gamma_{i,i+r}$ is the source of the equation or correlation functions can be written as
\begin{align}
    \label{eq:oslo_QQ(tt)_source}
    \Gamma_{i,i+r}(t) = \hata_i(t) \br{\delta_{i,{i+r}} - \delta_{i-1,{i+r}}}
    - \hata_{i+1}(t) \br{\delta_{i+1,{i+r}} - \delta_{i,{i+r}}}.
\end{align}
In the steady state, this source function can be written as
\begin{align}
    \label{eq:oslo_QQ(tt)_source_sdst}
    \Gamma_{r}(t) = a(\rhobar) \br{2\delta_{0,{r}} - \delta_{0,{r+1}} - \delta_{0,{r-1}}},
\end{align}
which is the derivation of Eq. \eqref{eq:oslo_gamma_r_sdst} in the main text.

\end{document}